\documentclass[11pt,a4paper]{article}
\usepackage{jcappub}
\usepackage{graphicx}
\usepackage{color}
\usepackage{amsfonts, amssymb,amsmath}
\usepackage{hyperref}
\usepackage{slashed}
\usepackage{braket}
\usepackage{comment}
\usepackage{xspace} 
\usepackage{siunitx}
\usepackage{multirow}
\usepackage{subcaption}
\usepackage{placeins}
\usepackage{afterpage}
\usepackage{array}
\usepackage{footmisc}
\usepackage[normalem]{ulem}
\usepackage[dvipsnames]{xcolor}
\usepackage{soul}

\newcolumntype{Z}{>{\setbox0=\hbox\bgroup}c<{\egroup}@{\hspace*{-\tabcolsep}}}

\usepackage{array}
\newcolumntype{P}[1]{>{\centering\arraybackslash}p{#1}}
\usepackage{multirow}

\definecolor{ForestGreen}{rgb}{0.13, 0.55, 0.13}
\definecolor{airforceblue}{rgb}{0.36, 0.54, 0.66}
\definecolor{orange}{rgb}{1.0, 0.5, 0.0}
\definecolor{amethyst}{rgb}{0.6, 0.4, 0.8}
\definecolor{awesome}{rgb}{1.0, 0.13, 0.32}
\definecolor{chromeyellow}{rgb}{1.0, 0.65, 0.0}

\newcommand{\quijote}{\textsc{Quijote}\xspace}
\newcommand{\quijotepng}{\textsc{Quijote-PNG}\xspace}
\newcommand{\sancho}{\textsc{Sancho}\xspace}

\begin{document}
\title{Cosmology with Persistent Homology: a Fisher Forecast}

\author[a]{Jacky H. T. Yip,}
\author[b,c,d]{Matteo Biagetti,}
\author[e,f,g]{Alex Cole,}
\author[f,g]{Karthik Viswanathan,}
\author[a]{and Gary Shiu.}

\affiliation[a]{Department of Physics, University of Wisconsin-Madison, Madison, WI 53706, USA}
\affiliation[b]{AREA Science Park, Padriciano 99, 34149 Trieste, Italy}
\affiliation[c]{Institute for Fundamental Physics of the Universe, Via Beirut 2, 34151 Trieste, Italy}
\affiliation[d]{Scuola Internazionale Superiore di Studi Avanzati, Via Bonomea 265, 34136 Trieste,  Italy}
\affiliation[e]{Gravitation Astroparticle Physics Amsterdam,
University of Amsterdam, Science Park 904, 1098 XH Amsterdam, The Netherlands}
\affiliation[f]{Delta Institute for Theoretical Physics, Science Park 904,
PO Box 94485, 1090 GL Amsterdam, The Netherlands}
\affiliation[g]{Institute of Physics, University of Amsterdam, Science Park
904, PO Box 94485, 1090 GL Amsterdam, The Netherlands}

\emailAdd{hyip2@wisc.edu}
\emailAdd{matteo.biagetti@areasciencepark.it}
\emailAdd{acole1221@gmail.com}
\emailAdd{k.viswanathan@uva.nl}
\emailAdd{shiu@physics.wisc.edu}

\abstract{Persistent homology naturally addresses the multi-scale topological characteristics of the large-scale structure as a distribution of clusters, loops, and voids. We apply this tool to the dark matter halo catalogs from the \quijote simulations, and build a summary statistic for comparison with the joint power spectrum and bispectrum statistic regarding their information content on cosmological parameters and primordial non-Gaussianity. Through a Fisher analysis, we find that constraints from persistent homology are tighter for $8$ out of the $10$ parameters by margins of $13-50\%$. The complementarity of the two statistics breaks parameter degeneracies, allowing for a further gain in constraining power when combined. We run a series of consistency checks to consolidate our results, and conclude that our findings motivate incorporating persistent homology into inference pipelines for cosmological survey data.}


\maketitle

\newpage

\section{Introduction}
\begin{figure}
    \centering
    \includegraphics[width=0.5\textwidth]{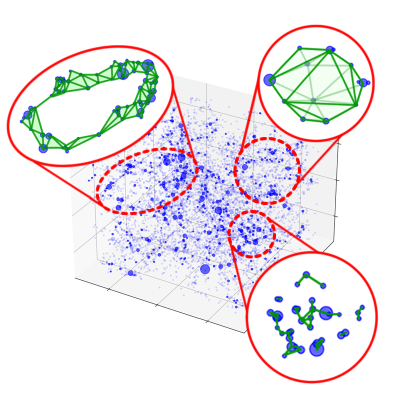}
    \caption{{\bfseries Topological data analysis of the large-scale structure.} Cosmic structures naturally correspond to topological features of various dimensions. Halo clusters (bottom right), filament loops (top left), and cosmic voids (top right) are $0$-, $1$-, and $2$-cycles in topology respectively.}
    \label{fig:conceptart}
\end{figure}

One of the ongoing challenges in modern precision cosmology is identifying and characterizing statistical summaries to extract the maximum amount of information from late-time observables. This task is complicated by the non-linearity of gravitational evolution, which takes initial fluctuations close to Gaussian to highly non-Gaussian structures. Therefore, optimal summary statistics are context-specific. Nevertheless, up to theoretical errors, methods of computing low-order correlation functions at large scales from perturbation theory techniques do provide access to some information. This approach has proven successful, particularly in the last few years, through the systematic analysis of BOSS data using predictions of the redshift space galaxy power spectrum and bispectrum up to one-loop order \cite{DAmico:2019fhj,Ivanov:2019pdj,Ivanov:2019hqk,Colas:2019ret,Zhang:2021yna,Cabass:2022wjy,DAmico:2022gki,Cabass:2022ymb,DAmico:2022osl,Ivanov:2023qzb}. These studies demonstrated the constraining power of galaxy surveys, providing competitive constraints as compared to those from the power spectrum of temperature anisotropies of the Cosmic Microwave Background (CMB) \cite{Aghanim:2018eyx,Akrami:2018odb,Planck:2019kim}. On the other hand, these statistics do not exhaust the information regarding our universe's initial conditions and dynamics. Moreover, statistics defined in Fourier space obscure consequences of position space locality, so that local perturbations to the data can have large effects. Most dramatically, some local effects (e.g., galaxy bias) cannot be distinguished from non-local physics (e.g., primordial non-Gaussianity) with these statistics (see \cite{Biagetti:2019bnp} for a review and references therein, and \cite{Baumann:2021ykm,Cabass:2023nyo} for a recent take on the topic).

Given these issues and because of the steady progress in computational techniques, a plethora of methods alternative to low-order correlation functions have been proposed in the last few years. A promising approach is that of field-level inference, which performs Bayesian inference of the initial conditions, as well as cosmological parameters, directly on voxels of the cosmic field \cite{Jasche:2012kq,Wang:2013ep,Ata:2014ssa,Schmittfull:2018yuk,Elsner:2019rql,Cabass:2019lqx,Cabass:2020nwf,Schmidt:2020viy,Cabass:2020jqo,Schmittfull:2020trd,Andrews:2022nvv,Bayer:2023rmj,Nguyen:2024yth}. The advantage of this approach is that it exploits, at least in principle, all the available information in a given map. The difficulty in its implementation is that it requires sampling a very high-dimensional parameter space: current generation survey volumes, such as Euclid and DESI, are of the order of $10\,($Gpc$/h)^3$ and resolve scales down to a few Mpc$/h$. This implies sampling in a space of $10^6-10^8$ dimensions, or more. Recent advances in samplers based on microcanonical Hamiltonian or Langevin Monte Carlo attempt to address this problem \cite{Robnik:2022bzs, Robnik:2023pgt,Bayer:2023rmj}. In a more general context, parameter estimation based on directly simulating the observables is a rapidly advancing field often referred to as simulation-based inference \cite{greenberg2019automatic,papamakarios2019sequential,hermans2020likelihood,Miller:2021hys,AnauMontel:2023stj,Heinrich:2023bmt} (see \cite{cranmer2020frontier} for a recent review and \cite{Alsing:2019xrx,Villaescusa-Navarro:2021cni,Villaescusa-Navarro:2021pkb,Makinen:2021nly,Cole:2021gwr,Lemos:2022kua,Hahn:2022wgo,Tucci:2023bag,List:2023jwo,Modi:2023llw,Modi:2023drt} for cosmological applications).

Instead of aiming at the full field-level information, one may construct by hand a suitable summary statistic, such that it contains more information than low-order correlation functions and retains a degree of physical interpretability of the observable. For inference, one can then build a neural network to map the statistic to parameters. The drawback of this is that irrespective of the neural network doing a perfect job, the summary itself is likely going to be, to a certain extent, a sub-optimal compression.

This approach can also be viewed as having a hard-coded layer for preprocessing the data before it is fed into an inference model, ensuring that only worthy information, which respects aspects such as symmetries, is used for inference. It is often desirable to use robust information against certain unknown physics, in other words, to use ``nuisance-hardened'' information \cite{Alsing:2019dvb}. It was advocated in \cite{Alsing:2019xrx} that in the context of cosmological inference, one performs a two-step compression process where the data is first summarized into well-understood hand-crafted statistics, and a neural network follows for inference, trained to extract maximum information from these summaries.

Hence, in recent years, there has been a surge in exploring new interpretable summary statistics that go beyond low-order correlations. Notable among these are marked fields \cite{Neyrinck:2009fs,White:2016yhs,Satpathy:2019nvo,Aviles:2019fli,Massara:2020pli,Philcox:2020srd,Philcox:2020fqx,Massara:2022zrf}, skew-spectra \cite{Schmittfull:2014tca,MoradinezhadDizgah:2019xun,Dai:2020adm,Schmittfull:2020hoi,Chakraborty:2022aok,Hou:2022rcd}, void statistics \cite{Pisani:2019cvo}, one-point PDFs \cite{Mao:2014caa,Uhlemann:2017tex,Nusser:2018vym,Friedrich:2019byw,Uhlemann:2019gni,Uhlemann:2022znd}, nearest neighbor distributions \cite{Banerjee:2020umh,Banerjee:2021hkg,Banerjee:2021cmi}, pair-connectedness functions \cite{Philcox:2022yri}, and wavelet scattering transforms \cite{Valogiannis:2021chp,Eickenberg:2022qvy,Valogiannis:2022xwu}---a list far from exhaustive.

Along these lines, several of the present authors have proposed in \cite{Biagetti:2020skr} an approach based on computational topology (see \cite{yip:2023} for an inference pipeline, and \cite{Cole:2017kve,Cole:2018emh} for earlier applications to the CMB and the string theory landscape). Namely, \cite{Biagetti:2020skr} constructed summary statistics from \emph{persistent homology} (see \cite{otter_roadmap_2017} for a review). The tool was used to quantify the large-scale structure morphology, via discrete and biased tracers of the matter field, as a distribution of clusters, loops, and voids across scales. Parametric tests were used to detect primordial non-Gaussianity of the local type in distributions of dark matter halos from N-body simulations, and the topological summaries were found to provide a level of sensitivity on par with the standard method of scale-dependent bias in the tracer's power spectrum. Furthermore, the topological signals were captured at mildly non-linear length scales, complementary to the long-wavelength modes relevant to scale-dependent bias. This indicates promising potential of the large-scale structure's topology for alternative non-Gaussianities. Of this sort, in \cite{Biagetti:2022qjl}, lower-dimensional summaries enabled the calculation of (the inverse of) the covariance matrix and the Fisher information for the local and equilateral types of primordial non-Gaussianity, also showing promising results. In parallel, similar methods have been applied to matter fields \cite{Kanafi:2023twi} and weak lensing maps \cite{Heydenreich:2022dci,Heydenreich:2020hrr} to derive parameter constraints.


One good aspect of taking this approach is that statistics derived from this tool are naturally functions of distance, and may thus be recognized as syntheses of information from a range of high-order correlators ``filtered'' through topology. This is evident when considering that each topological feature originates from varying numbers of halos as vertices, which we will explain later. To this end, persistent homology for cosmology deserves a more thorough analysis. In this work, we further expand the power of the previous setups by varying the nearest-neighbor parameter $k$ in the coarse-graining scheme to emphasize topological features of different scales. Then we apply our pipeline to the \quijote halo catalogs \cite{Villaescusa-Navarro:2019bje, Coulton:2022qbc,Coulton:2022rir}, and compute constraints via the Fisher information matrix for $7$ cosmological parameters and $3$ primordial non-Gaussianity amplitudes.

\begin{figure}
    \centering
    \includegraphics[width=1\textwidth]{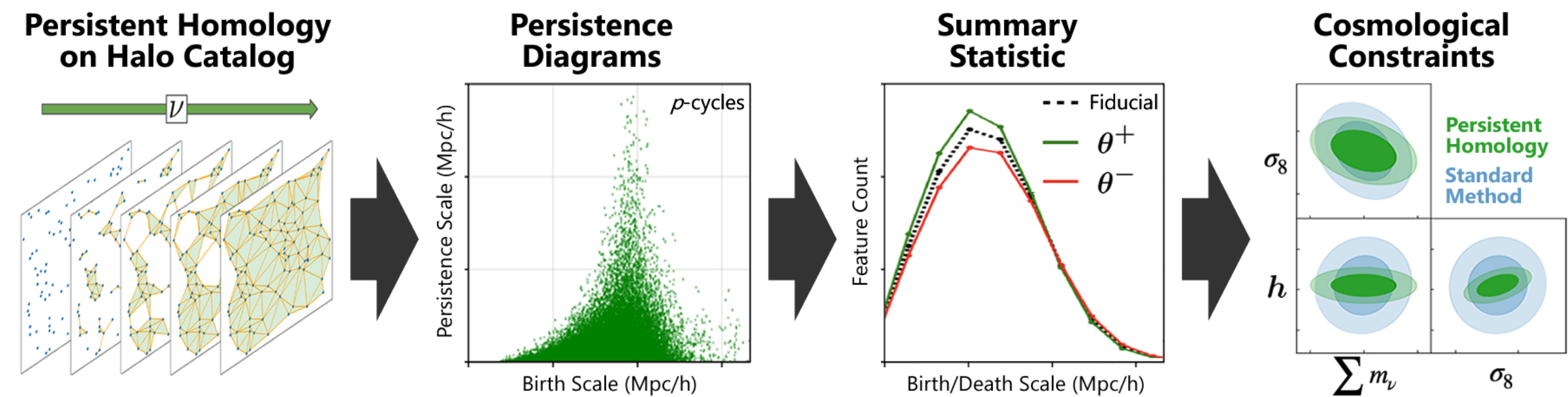}
    \caption{{\bfseries Flow of this paper.} We explain the machinery of persistent homology and detail the filtration designed for dark matter halos. The raw outputs of a persistent homology analysis are persistence diagrams, which we summarize into a histogram-based statistic. Applying this pipeline to the entire suite of halo catalogs, we compute the Fisher information matrix and report our cosmological constraints, which are generally tighter in comparison with the joint power spectrum and bispectrum statistic.}
    \label{fig:pop}
\end{figure}

The flow of this paper is presented in Figure \ref{fig:pop}. Section \ref{sec:dataset} introduces the dataset, specifically the dark matter halo catalogs from \quijote in redshift space. In Section \ref{sec:PHreview}, we review the essential aspects of persistent homology, with a focus on the tailored coarse-graining scheme for halos. In Section \ref{sec:summarystat}, we describe the building of our summary statistic from the outputs of a persistent homology computation. The results of our Fisher analysis are presented in Section \ref{sec:results}, where we discuss them in comparison with conventional statistics of low-order correlation functions. Finally in Section \ref{sec:conclusions}, we give our conclusions and outlook for future research directions.

\section{Dataset}\label{sec:dataset}
Our method of persistent homology is well-defined for any point cloud, such as positions of dark matter halos or galaxies. In this work, we apply our pipeline to dark matter halos. The halo catalogs at redshift $z=0.5$ in real space are taken from the \quijote and \quijotepng suites \cite{Villaescusa-Navarro:2019bje,Coulton:2022qbc} (hereafter collectively referred to as \quijote). The choice of $z=0.5$ is relatable to galaxies observed in surveys such as the BOSS CMASS sample \cite{SDSS-III:2015hof,Reid:2015gra}.\footnote{We have tried with considerable efforts to use the \sancho mock galaxy dataset, previously used in \cite{Gambardella:2023qrm}, which is modeled to mock CMASS galaxies. However, we found that key assumptions necessary for the Fisher analysis break down for this dataset. See Appendix \ref{app:sancho} for details.} The \quijote N-body simulations are run with \texttt{GADGET-III}, each having $512^3$ dark matter particles in a $1\,($Gpc$/h)^3$ box, and the halo catalogs are generated from snapshots with the friends-of-friends algorithm. We use halos that are composed of at least $50$ particles, which corresponds to a minimum halo mass of $M_{\rm min}=3.28\times10^{13}M_{\odot}$ and an average halo number density of $\bar{n}\approx9.88\times10^{-5}\,(h/{\rm Mpc})^3$ at the fiducial cosmology.

The halo catalogs, originally available in real space, are converted to catalogs in redshift space using the distant-observer approximation. We apply
\begin{equation}    \boldsymbol{x}\rightarrow\boldsymbol{x}+\frac{\boldsymbol{v}\cdot\hat{\boldsymbol{n}}}{a(z)H(z)}\hat{\boldsymbol{n}}
\end{equation}
to each halo position $\boldsymbol{x}$, where $\boldsymbol{v}$ is the halo's peculiar velocity, $\hat{\boldsymbol{n}}$ is the line-of-sight, $a(z)=(1+z)^{-1}$ is the scale factor, and the Hubble parameter $H(z)=100\sqrt{\Omega_{\rm m}a^{-3}+(1-\Omega_{\rm m})a^{-3(1+w)}}$ in $({\rm km}/{\rm s})(h/{\rm Mpc})$ depends on the specific cosmology of the catalog. We process each catalog along each of the three axes (z-axis only for the fiducial catalogs), following \cite{Hahn:2019zob}. Hence, there are effectively ($500$ N-body) $\times$ ($3$ RSD axes) $=1500$ realizations for each non-fiducial cosmology and $14500$\footnote{We skip the first $500$ of the $15000$ fiducial simulations available in \quijote as they share the same initial random seeds with the non-fiducial simulations. This ensures that the covariance and derivative estimations are completely independent.} realizations for the fiducial cosmology (Table \ref{tab:sachoSum}).\footnote{Hereafter we use the terms ``catalog'', ``simulation'', and ``realization'' interchangeably; they all refer to a simulated set of halo positions in redshift space.}

The fiducial cosmological parameters\footnote{The total mass of neutrinos $\sum m_\nu$ and the dark energy equation of state parameter $w$ are nominally included in what we refer to as the set of cosmological parameters.} are $\Omega_{\rm{m}}=0.3175$, $\Omega_{\rm{b}}=0.049$, $h=0.6711$, $n_{\rm{s}}=0.9624$, $\sigma_8=0.834$, $\sum m_\nu = 0.0\,$eV, and $w=-1$. In the non-fiducial cosmologies\footnote{Contrary to \quijote, we define the $w^+$ cosmology as the one with the less negative $w$ value.}, these values are varied by a step $\Delta\theta$ above and below:
\begin{align*}
    \{\Delta\Omega_{\rm{m}}, \Delta\Omega_{\rm{b}}, \Delta h, \Delta n_{\rm{s}}, \Delta \sigma_8,\Delta w\} = \{0.01, 0.002, 0.02, 0.02, 0.015,0.05\}.
\end{align*}
As for the total mass of neutrinos, $\sum m_\nu$, there are three non-fiducial values of $0.1$, $0.2$, and $0.4\,$eV.

The realizations from \quijotepng have initial conditions with different shapes of primordial non-Gaussianity: local \cite{Salopek:1990jq}, equilateral \cite{Creminelli:2005hu}, and orthogonal-LSS \cite{Senatore:2009gt} (these shapes are special cases of the bispectrum for general single field inflation computed in \cite{Chen:2006nt}). The steps for them are all $\Delta f_{\rm NL} = 100$.

\begin{table}[t]
\scriptsize
\begin{centering}
\begin{tabular}{ |P{2cm}|P{3.75cm}|P{3.75cm}|P{3.75cm}| } 
    \hline
    \textbf{N-body suite} & \textbf{Cosmology} & \textbf{Varying parameter value} & \textbf{Number of realizations}\\
    \hline
    \hline
    \multirow{17}*{\quijote}
    & Fiducial & - & $14500$\\ 
    \cline{2-4}
    & $\Omega_{\rm m}^+$ & $0.3275$ & $1500$\\
    \cline{2-4}
    & $\Omega_{\rm m}^-$ & $0.3075$ & $1500$\\
    \cline{2-4}
    & $\Omega_{\rm b}^+$ & $0.051$ & $1500$\\
    \cline{2-4}
    & $\Omega_{\rm b}^-$ & $0.047$ & $1500$\\
    \cline{2-4}
    & $h^+$ & $0.6911$ & $1500$\\
    \cline{2-4}
    & $h^-$ & $0.6511$ & $1500$\\
    \cline{2-4}
    & $n_{\rm s}^+$ & $0.9824$ & $1500$\\
    \cline{2-4}
    & $n_{\rm s}^-$ & $0.9424$ & $1500$\\
    \cline{2-4}
    & $\sigma_{8}^+$ & $0.849$ & $1500$\\
    \cline{2-4}
    & $\sigma_{8}^-$ & $0.819$ & $1500$\\
    \cline{2-4}
    & $(\sum m_\nu)^{+++}$ & $0.4$ & $1500$\\
    \cline{2-4}
    & $(\sum m_\nu)^{++}$ & $0.2$ & $1500$\\
    \cline{2-4}
    & $(\sum m_\nu)^+$ & $0.1$ & $1500$\\
    \cline{2-4}
    & Fiducial (Zeldovich IC) & - & $1500$\\
    \cline{2-4}
    & $w^+$& $-0.95$ & $1500$\\
    \cline{2-4}
    & $w^-$ & $-1.05$ & $1500$\\
    \hline
    \hline
    \multirow{6}*{\quijotepng} 
    & $f_{\rm NL}^{\rm{local}+}$ & $100$ & $1500$\\
    \cline{2-4}
    & $f_{\rm NL}^{\rm{local}-}$ & $-100$ & $1500$\\
    \cline{2-4}
    & $f_{\rm NL}^{\rm{equil}+}$ & $100$ & $1500$\\
    \cline{2-4}
    & $f_{\rm NL}^{\rm{equil}-}$ & $-100$ & $1500$\\
    \cline{2-4}
    & $f_{\rm NL}^{\rm{ortho}+}$ & $100$ & $1500$\\
    \cline{2-4}
    & $f_{\rm NL}^{\rm{ortho}-}$ & $-100$ & $1500$\\
    \hline
\end{tabular}
\begin{tabular}{ P{2.03cm}P{3.75cm}|P{3.75cm}|P{3.75cm}| }
    \cline{3-4}
    &  & \textbf{Total} & $47500$ \\
    \cline{3-4}
\end{tabular}
\par\end{centering}
\caption{\label{tab:sachoSum} \textbf{Organization of all realizations used.}}
\end{table}

\paragraph{Power spectrum and bispectrum measurements.} We measure the halo auto power spectrum monopole and quadrupole ($P_{\ell=0}$ and $P_{\ell=2}$), and the bispectrum monopole ($B_{\ell=0}$), using the public code \textsf{PBI4} \cite{biagetti_2023_10008045}. The code uses a fourth-order density interpolation and interlacing scheme as described in \cite{Sefusatti:2015aex}. Bins have a width of $\Delta k = 2 k_f$, where $k_f=0.006\,h/$Mpc is the fundamental frequency of the $1\,($Gpc$/h)^3$ box, and measurements are taken up to $k_{\rm max} = 0.3\,h/$Mpc, beyond which 
the shot-noise, assumed to be Poissonian\footnote{While this is a standard assumption, the actual shot-noise might tend to be sub-Poissonian as the halos used in this work fall into the higher-mass end. The authors thank the Referee for pointing this out.}, strongly dominates\footnote{The signal-to-noise ratio drops below $0.5$ beyond $k_{\rm max}=0.3\,h$/Mpc for $P_0$.}. This results in $24$ bins in each of $P_0$ and $P_2$, and $1522$ triangles for $B_0$. We subtract shot-noise from the monopole measurements, except for the fiducial measurements which are used for estimating the covariance. The Poisson shot-noise terms are $1/\bar{n}$ for $P_0$ and $\frac{1}{\bar{n}^2}+\frac{P_0(k_1)+P_0(k_2)+P_0(k_3)}{\bar{n}}$ for $B_0(k_1, k_2, k_3)$. Since the focus of this work is not on these conventional low-order correlation statistics, we do not show further details about these measurements; the reader may refer to \cite{Hahn:2019zob,Coulton:2022rir}.

\section{From Dark Matter Halos to Persistence Diagrams}\label{sec:PHreview}
At large scales, dark matter is organized in a cosmic web of clusters, filaments, and voids of different sizes and shapes \cite{Bond:1995yt}. The goal of our method is to characterize these cosmic structures by applying persistent homology to dark matter halo positions, then quantify the information content on parameters of the $\Lambda$CDM cosmological model and primordial non-Gaussianity.

The main concepts of a persistent homology computation can be summarized in simple words as follows. From a \emph{point cloud} of halos, we build a series of \emph{simplicial complexes} by adding simplices one after another according to their length scale. Each \emph{topological feature}, as a ``hole'' in the series of complexes, only exists and ``persists'' over a continuous range of scales, since the simplices that assemble or destroy the feature have not been added at the two scales that define this range. We then collect this data from all topological features and construct a statistical summary informative for cosmology.

In this section, after reviewing basic topology, we explain the process described above in detail, in particular how diagrams of topological features are built from a halo catalog. While in this paper we work with simulated halos in a cubic box that might not be observationally realistic, we emphasize that our pipeline is also well-defined for actual data, e.g., galaxies in a lightcone with a given survey geometry and window function. Nevertheless, this work is a proof of concept that reveals the amount of information statistics from persistent homology can contain for cosmology, which serves as the basis for further developing parameter inference pipelines.

\subsection{Coarse-graining}\label{subsec:cograin}
\begin{figure}
    \centering
    \includegraphics[width=1\textwidth]{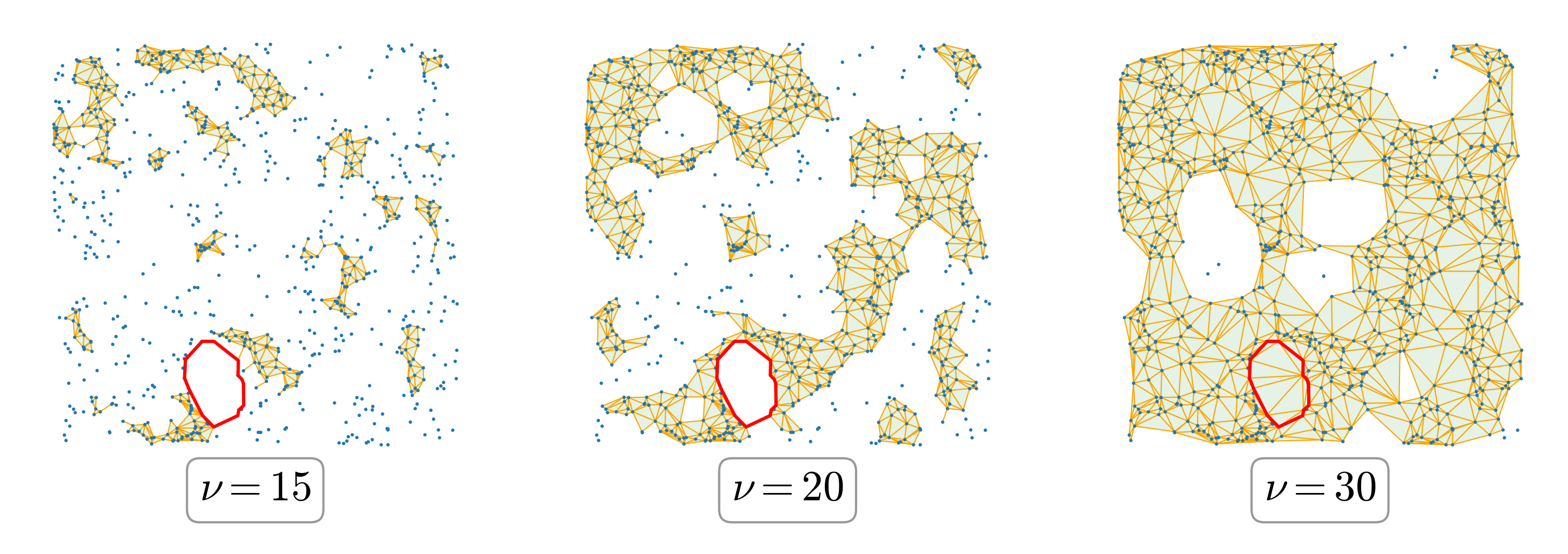}
    \caption{{\bfseries Life of a topological feature.} The blue point cloud represents projected redshift space positions of halos in a $300\times 300\times 100\,$(Mpc$/h$)$^3$ subbox within a fiducial realization. We illustrate the ``life'' of an underdensity (enclosed by the red boundary) as a topological feature in the evolving simplicial complex as the filtration parameter $\nu$ grows (analogously the coarse-graining scale). \emph{Left:} At a small scale, the simplicial complex contains none of the simplices that mark out the underdensity. \emph{Middle:} At an intermediate scale, all the relevant edges have just been added. Born as a ``hole'' in the complex, the underdensity can be associated to a topological feature, a member of a homology class of $1$-cycles. \emph{Right:} At a later stage, the hole is completely filled with triangular faces---the feature has died.}
    \label{fig:cartoonDTM}
\end{figure}

Before defining the kind of operations we would like to perform on a set of halos based on topology, we build the cosmologist's intuition through coarse-graining concepts. As noted above, the large-scale structure of the universe is a complex hierarchy of clusters, filament loops, and cosmic voids across length scales. These physical objects naturally correspond to the features that arise from a topological computation on the $3$-dimensional spatial distribution of halos (Figure \ref{fig:conceptart}). The multi-scale nature of the distribution implies that a feature may be distinguishable only within a restricted range of coarse-graining scales.\footnote{At this point we rely on an intuitive picture of coarse-graining. We define specific coarse-graining schemes (filtrations) in Section \ref{sec:cgvf}.} For example, two clusters separated by a distance $L$ will no longer be distinct once we coarse-grain to a length scale greater than $L$. Similarly, a cosmic void only appears when the tracers defining its boundary are connected, and it will diminish and ultimately vanish when its interior fills up as the length scale increases further. This interplay between the coarse-graining scale and the births and deaths of features is illustrated in Figure\ \ref{fig:cartoonDTM}.

Our approach in this work is to use persistent homology to track the distribution of topological features across length scales. The formalism of persistent homology lets us compute the statistics of the scales at which these features are created and destroyed (and the difference between these two, i.e., a given feature's ``persistence''). 

\subsection{Topology of simplicial complexes}
Topology investigates the set of properties of a mathematical space that are preserved under continuous deformations. To define a topological space for data represented as a point cloud, i.e., a set of discrete points, we seek a way of linking the points, embedding the data within a simplicial complex. A simplicial complex is an assembly of $n$-simplices which, in our $3$-dimensional context, includes vertices ($n=0$), edges ($n=1$), triangles ($n=2$), and tetrahedra ($n=3$). For instance, consider four points where point A and point C are connected by a line segment, and point B and D are disconnected. We may define a simplicial complex as the set that includes all the simplices, i.e., the four vertices (A, B, C and D) and the one edge, AC. Subsets of this set are also simplicial (sub)complexes, e.g., the subset that contains only the four vertices. These sets are built to be closed under the intersection of simplices and under taking faces (e.g., if an edge is present, the two vertices making up its boundary are also present).

\begin{figure}
    \centering
    \includegraphics[trim={2cm 2cm 2cm 2cm},clip,width=0.40\textwidth]{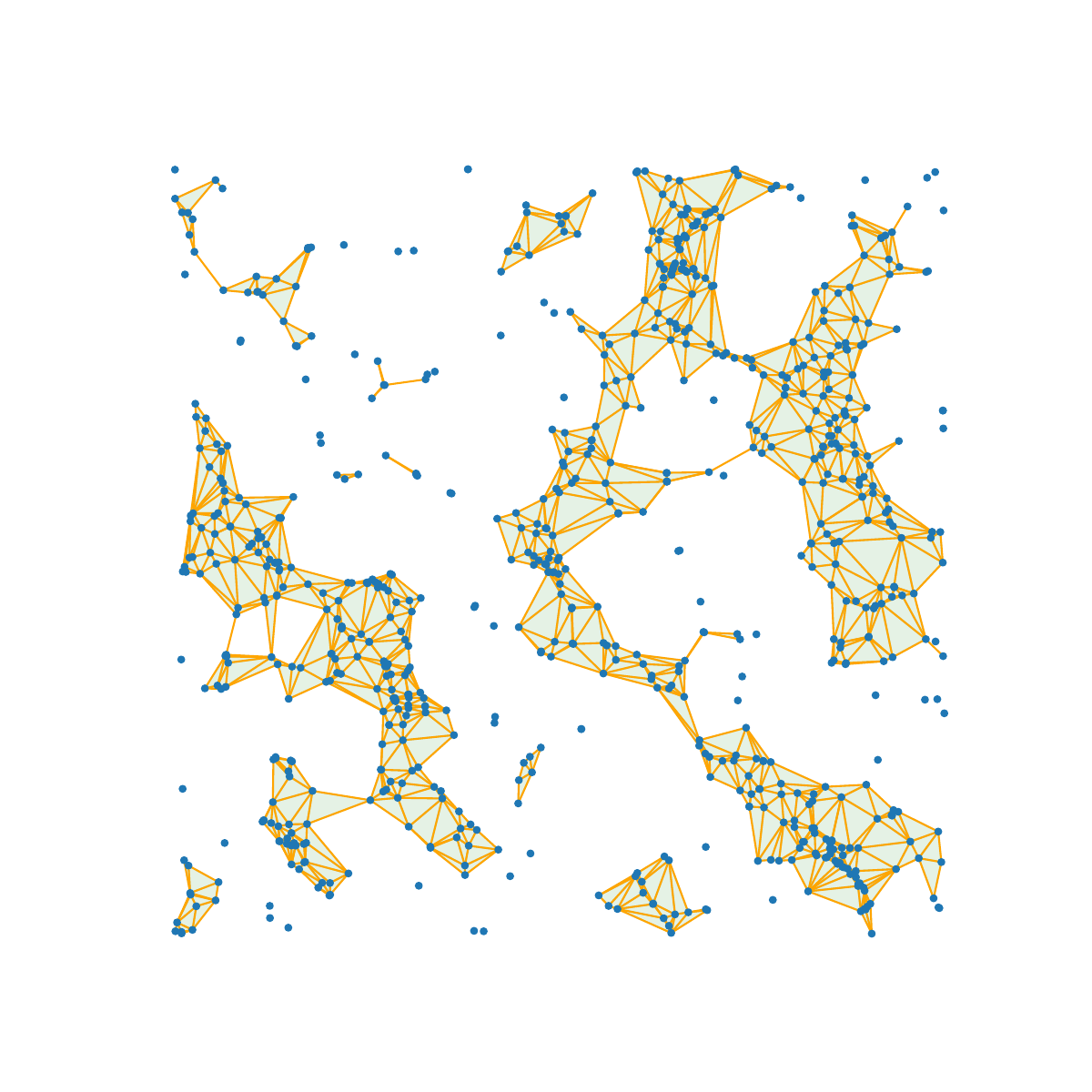}
    \caption{\textbf{Emergence of topological features.} A filtration is a nested family of simplicial complexes. Shown here is the simplicial complex at a particular filtration time in the $\alpha$-DTM$\ell$-filtration of a point cloud (here we ignore the fact that vertices have non-trivial birth times and plot all of them for illustrative purposes). Vertices, edges, triangles, and tetrahedra ($n$-simplices) from the Delaunay triangulation of the point cloud are gradually added to the complex, at designated filtration times determined by their sizes and the distributions of their surrounding neighbors. Topological features as $(p+1)$-dimensional holes, formally known as equivalence classes of $p$-cycles, emerge and vanish throughout the filtration.}
    \label{fig:introcartoon}
\end{figure}

Given a simplicial complex built on a point cloud, we would like to identify deformation-invariant objects. This is done by defining linear operators $\partial_p$ that take a collection of $p$-simplices to its $(p-1)$-dimensional boundary. In words, a topological feature corresponds to a collection of $p$-simplices with vanishing boundaries that are not themselves the boundary of a collection of $(p+1)$-simplices. The identification of a feature is done by finding the boundary of appropriate combinations of $(p+1)$-simplices. In particular, it is useful to define the quotient group $H_p \equiv \ker \partial_p/\textrm{im}\partial_{p+1}$, i.e., the $p$-th homology group of the complex. Each element of $H_p$ is an independent ``hole'' of dimension $p$. More formally, the elements of $H_p$ are \emph{equivalence classes of $p$-cycles}---collections of $p$-simplices with vanishing boundaries.\footnote{In plain language, a $0$-cycle is a collection of connected simplices, a $1$-cycle is a loop, and a $2$-cycle is an enclosed cavity.} The intuitive connection to deformation-invariance is that two cycles belonging to the same homology class, as an equivalence class, can deform into each other. Consequently, cycles that envelop distinct holes cannot be in the same homology class. As an explicit example, there are only three homology classes of $1$-cycles in Figure \ref{fig:introcartoon} (two small holes adjacent to each other on the left, a bigger one on the right), while one may identify many closed paths that circle each of them. For clarity, we point out that the addition of a $p$-simplex leads to either the birth of a new homology class of $p$-cycles, or the death of an existing homology class of $(p-1)$-cycles. We refer the reader to \cite{hatcher,Dey_Wang_2022} for more mathematical details.

\subsection{Coarse-graining via filtration}\label{sec:cgvf}
A single simplicial complex is insufficient to encapsulate the step-by-step details of the coarse-graining procedure described earlier in Section \ref{subsec:cograin}. A family of complexes is required to capture the evolving topology through the coarse-graining scales. Such a family is called a \emph{filtration}, with the coarse-graining scale represented by the non-negative filtration parameter, also called the \emph{filtration time}, $\nu$. The filtration parameter can be used to track the time $\nu_{\text{birth}}$ at which a homology class is born, and the time $\nu_{\text{death}}$ at which it dies later at a larger scale. The persistence $\nu_{\text{persist}}$ is the difference between the two, i.e., $\nu_{\text{persist}}\equiv\nu_{\text{death}}-\nu_{\text{birth}}$.

An intuitive example of a filtration is the Vietoris-Rips (VR) filtration. Let us consider a generic point cloud $X$ embedded in a $N$-dimensional metric space equipped with a distance function $d(x,y)$. The set of vertices of each simplicial complex in this filtration is the point cloud itself,  $X$. The criterion to include an edge connecting two vertices is that their separation satisfies $d\leq2\nu$. The intuitive picture here is that we can draw a growing $N$-ball of radius $\nu$ around each vertex. Any time when, or after, two balls touch, the associated edge is included in the simplical complex. Higher-dimensional simplices are included if all necessary faces are present. Although intuitive, the VR filtration does not suit our point clouds because the inclusion scheme is rather inefficient: as $\nu \to \infty$ there is an edge for every pair of vertices, a triangle for every triplet of vertices, and so on. The computational cost becomes prohibitively expensive for large datasets like ours. In the following, we focus on filtrations that are similar in spirit to the VR filtration, but computationally more efficient.

\paragraph{$\alpha$-filtration.}
Our method is based on the intuitive $\alpha$-filtration \cite{edelsbrunner1994three}. In contrast with the VR filtration, the embedding of the point cloud is done via its \emph{Delaunay triangulation} rather than to universally allow for an edge between any two vertices. Then, the simplicial complexes at different times in an $\alpha$-filtration are subcomplexes of the Delaunay complex \cite{edelsbrunner1994three}.\footnote{The Delaunay complex can be defined in the following way. Given a set of points $X\subset \mathbb{R}^d$, the \emph{Voronoi cell} of a point $x\in X$ is defined as $V_x=\{y\in \mathbb{R}^d ~|~ ||x-y||\leq ||z-y||~\forall z\in X\}$. Informally, $V_x$ includes all points in $\mathbb{R}^d$ to which $x$ is the nearest (or tied as the nearest) among all other points in $X$. The collection of these cells triangulates the space in which the point cloud is embedded and constitutes the \emph{Voronoi diagram of $X$}. The Delaunay complex is defined via the Voronoi diagram by $\{\sigma\subseteq X~|~\cap_{x\in \sigma} V_x\neq \emptyset\}$.} The appealing aspect of Delaunay subcomplexes is the dramatic reduction in the total number of simplices. This reduction offsets the computational overhead involved in finding these complexes, making the overall algorithm more computationally efficient than the VR filtration.\footnote{This is related to the fact that the Delaunay triangulation is inherently geometric, as opposed to the VR complex in which the number of simplices is very large, yet many are ``unphysical''.}

The fact that every point is treated equally in the $\alpha$-filtration, in the sense that it makes no difference to the algorithm whether or not a point is in a sparsely populated region, is perhaps an unappealing property. As an example, consider a large, empty void that dies at $\nu'$. If we put a point in its center, the feature will die at $\nu'/2$ instead. This is not desirable as this single point should be considered as an outlier, not impacting significantly the defining length scale of the void. That is to say, the $\alpha$-filtration is not stable against the addition of outliers. In light of this, we introduce the $\alpha$-DTM$\ell$-filtration.\footnote{Note that in a cosmological sense, no halo or galaxy is an ``outlier''. Nevertheless, as we show below, introducing a weight on each point based on its neighbors effectively helps extract information from different scales. In fact, it is reasonable to expect that topological signals of interesting physics may be recovered more optimally by tuning a local bias-like parameter. For example, underdense features like voids are sensitive to diffuse sources such as massive neutrinos \cite{Zhang:2019wtu}. Moreover, it has been shown that it might be interesting to select outlier galaxies inside voids \cite{Wang:2024dvm}. Our approach can be modified to investigate these ideas in more detail; these studies are reserved for future work.}

\begin{figure}
    \centering
    \includegraphics[scale=0.22]{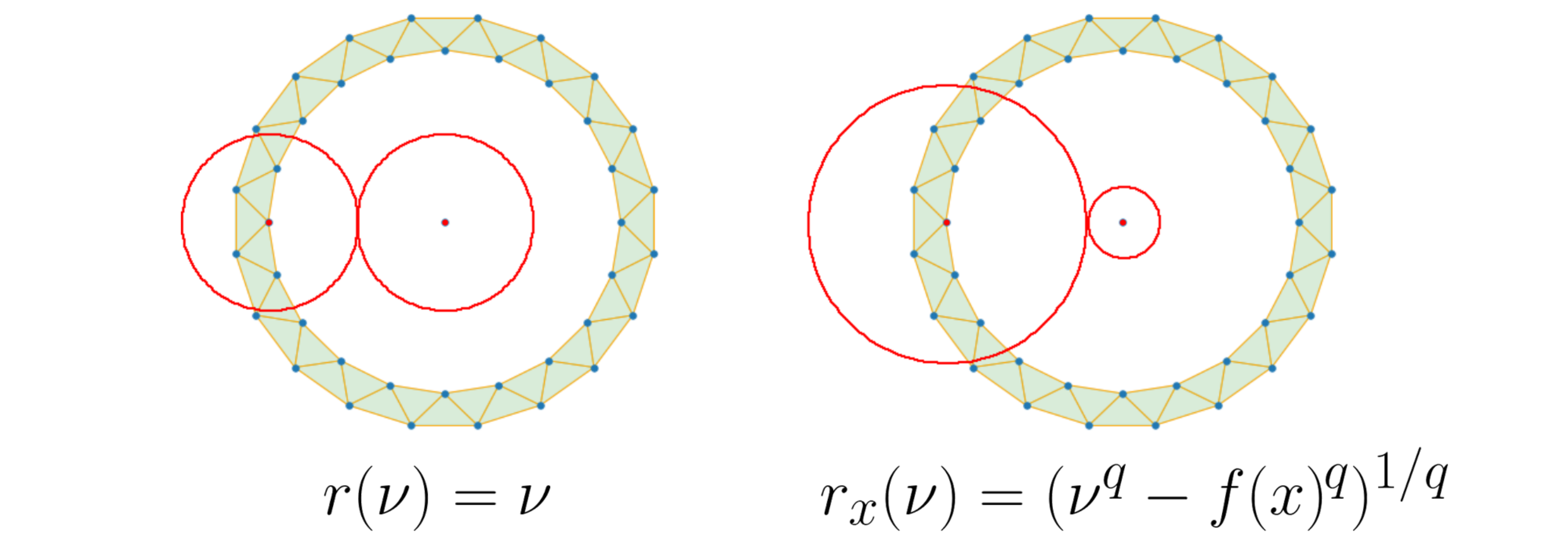}
    \caption{\textbf{$\alpha$-filtration vs. $\alpha$-DTM$\ell$-filtration.} Here we use as an example a $1$-cycle (the circular loop) with an outlier (a point) in the center. \emph{Left:} The growing ball around every point has a uniform radius that is equal to the filtration time $\nu$. \emph{Right:} Characterized by the Distance-to-Measure function $f$, the more outlying a point is, the more impeded the growth of its ball is. Therefore, edges and higher-dimensional simplices that trivialize this $1$-cycle are added at postponed times, i.e., the $1$-cycle survives longer. Hence, the $\alpha$-DTM$\ell$-filtration is more stable against the addition of outliers.}
    \label{fig:outlierdemo}
\end{figure}

\paragraph{$\alpha$-DTM$\ell$-filtration.}To lessen this disproportionate effect of outliers, the filtration time at which a simplex is added should be postponed depending on how much of an outlier each of the vertices involved is. Recall that, in the VR filtration, a $1$-simplex is added when the imaginary balls around two vertices touch. In the $\alpha$-DTM$\ell$-filtration \cite{chazal2017robust}, instead of using $\nu$ as the uniform radius, we employ the point-specific radius function
\begin{equation}
\label{rx}
    r_x(\nu)=(\nu^q-f(x)^q)^{1/q},
\end{equation}
i.e., the radius of the ball around point $x$ at a given $\nu$ depends on a function $f$ of the point itself, and a parameter $q$ that controls the mixing between $\nu$ and $f(x)$. The parameter $q$ could be made tunable to investigate how to maximize (minimized) certain desired (undesired) aspects of the filtration. Here we choose $q=2$ and devote a thorough investigation into an optimal choice to future work. Note that $f(x)$ accounts for the ``unimportance'' of the point $x$ and impedes the growth of its ball as shown in Figure \ref{fig:outlierdemo}. Hence, $f(x)$ should take a large value for a halo that lives in a sparsely populated region, and vice versa. The \emph{Distance-to-Measure} (DTM) function suits this purpose:
\begin{equation}\label{eq:dtm}
    f(x)=\text{DTM}(x)\equiv\left(\frac{1}{k}\sum\limits_{X_i\in \mathcal{N}_k(x)}\|x-X_i\|^p\right)^{1/p}
\end{equation}
where $\mathcal{N}_k(x)$ is the set of $k$-nearest neighbors to $x$, and $p$ is a mixing parameter. For Eq. \eqref{rx} to properly describe a radius, any point $x$ is added to the simplicial complex at $\nu=f(x)$. As a result, $\nu_\text{birth}$ values of $0$-cycles become nontrivial and contain information on the local density of the point cloud. As before, high-dimensional simplices are added when all faces are present. 

\begin{figure}
    \centering
    \includegraphics[width=1\textwidth]{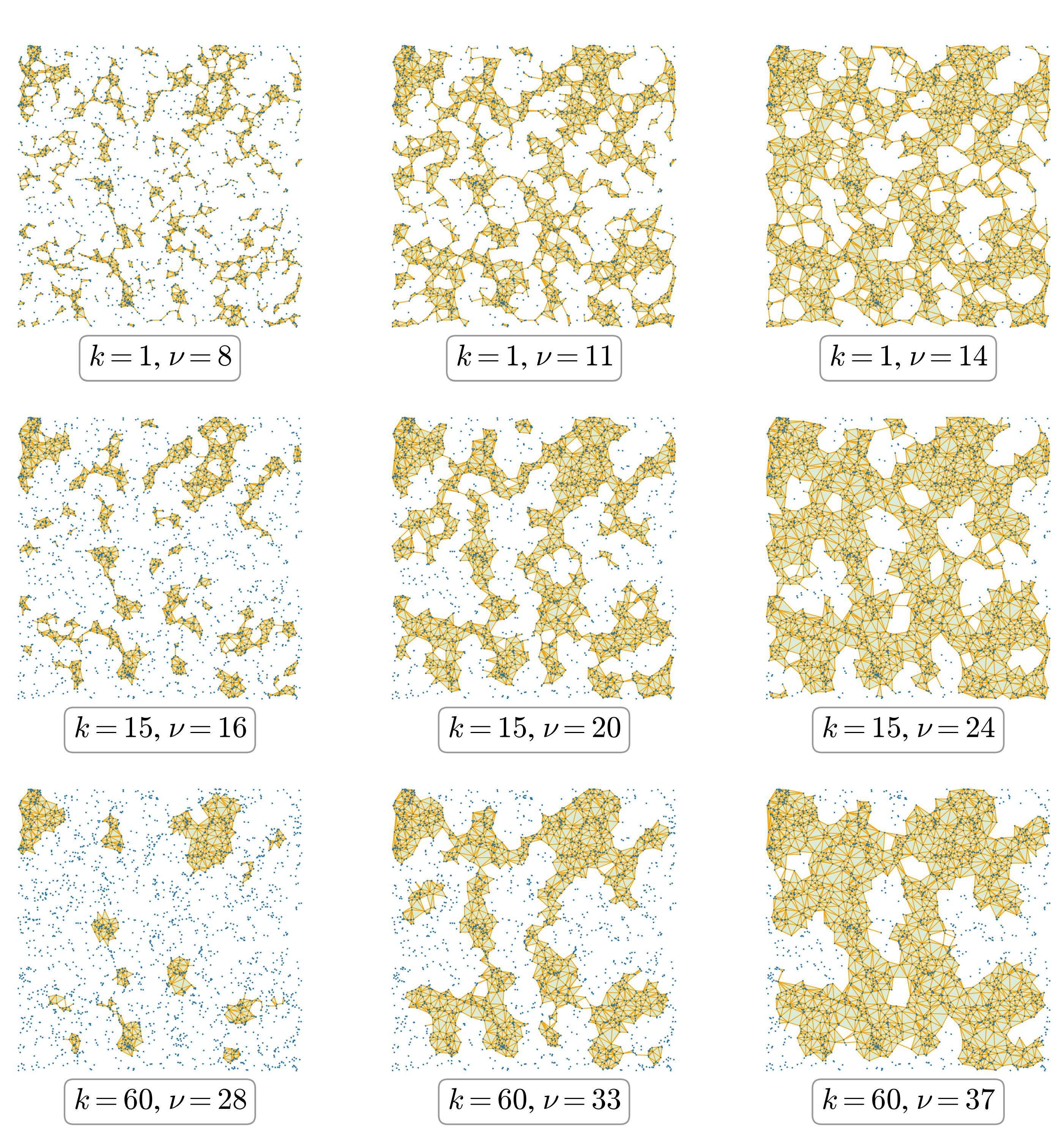}
    \caption{\textbf{Effect of $k$ on the $\alpha$-DTM$\ell$-filtration.} \emph{Left to Right:} Early, middle, and late stages in the $\alpha$-DTM$\ell$-filtration of a sample of \quijote halos (for clarity, all halos are plotted regardless of their birth times as $0$-cycles). \emph{Top to Bottom:} Increasing value of the nearest-neighbor parameter $k$. For a fair comparison between $k$ values in each column, the filtration time $\nu$ at which we inspect the simplicial complex varies with $k$, since birth (and death) times of features are positively offset when a larger $k$ is used. Across all stages, there are fewer small-scale $0$- and $1$-cycles in the larger-$k$ setups. This suggests that a large-$k$ filtration encompasses the persistent homology of large-scale topological features, and vice versa. We encourage the reader to watch the filtrations in action: \url{https://youtu.be/_phgkiZmY0c}.}
    \label{fig:kEffect}
\end{figure}

Heuristically, $\text{DTM}(x)$ calculates the average distance from $x$ to
its $k$-nearest neighbors when $p=1$. Again, $k$ and $p$ are tunable parameters in principle. While we choose $p=2$ in this work such that DTM$(x)$ is the root mean square distance, it turns out that the nearest-neighbor parameter $k$ significantly impacts the filtration. In Figure \ref{fig:kEffect}, we show snapshots of the simplicial complex at different times (early to late from left to right) in the filtration for $k=1$, $15$, and $60$ (top to bottom). Note that to fairly compare $k$ values (snapshots in the same column), we must also vary the filtration time $\nu$. This is because $f(x)$ which controls the birth times of vertices increases monotonically with $k$, then in general birth and death times of all topological features are positively offset when a larger $k$ is used. From the comparison, we observe that fewer small-scale features emerge across all stages in the filtration for a larger $k$, suggesting that a large-$k$ filtration encompasses the persistent homology of large-scale features. This is no surprise, as increasing $k$ effectively increases the average volume within which the algorithm explores for the densest regions to populate with simplices. Hence, when $k$ is set to be larger, some regions are deemed no longer dense enough for early population, leaving space for larger holes (features) to emerge.

Given a single value of $k$, whether the features tracked are considered large depends on the volume and the number density of the point cloud. On this account, we vary $k$ and make use of all the resulting filtrations to maximize the amount of topological information we extract. This is done by concatenating the data vectors, each from one filtration, which we discuss in the following sections.

\subsection{Persistence diagrams and images}
\begin{figure}
    \centering
         \includegraphics[width=1\textwidth]{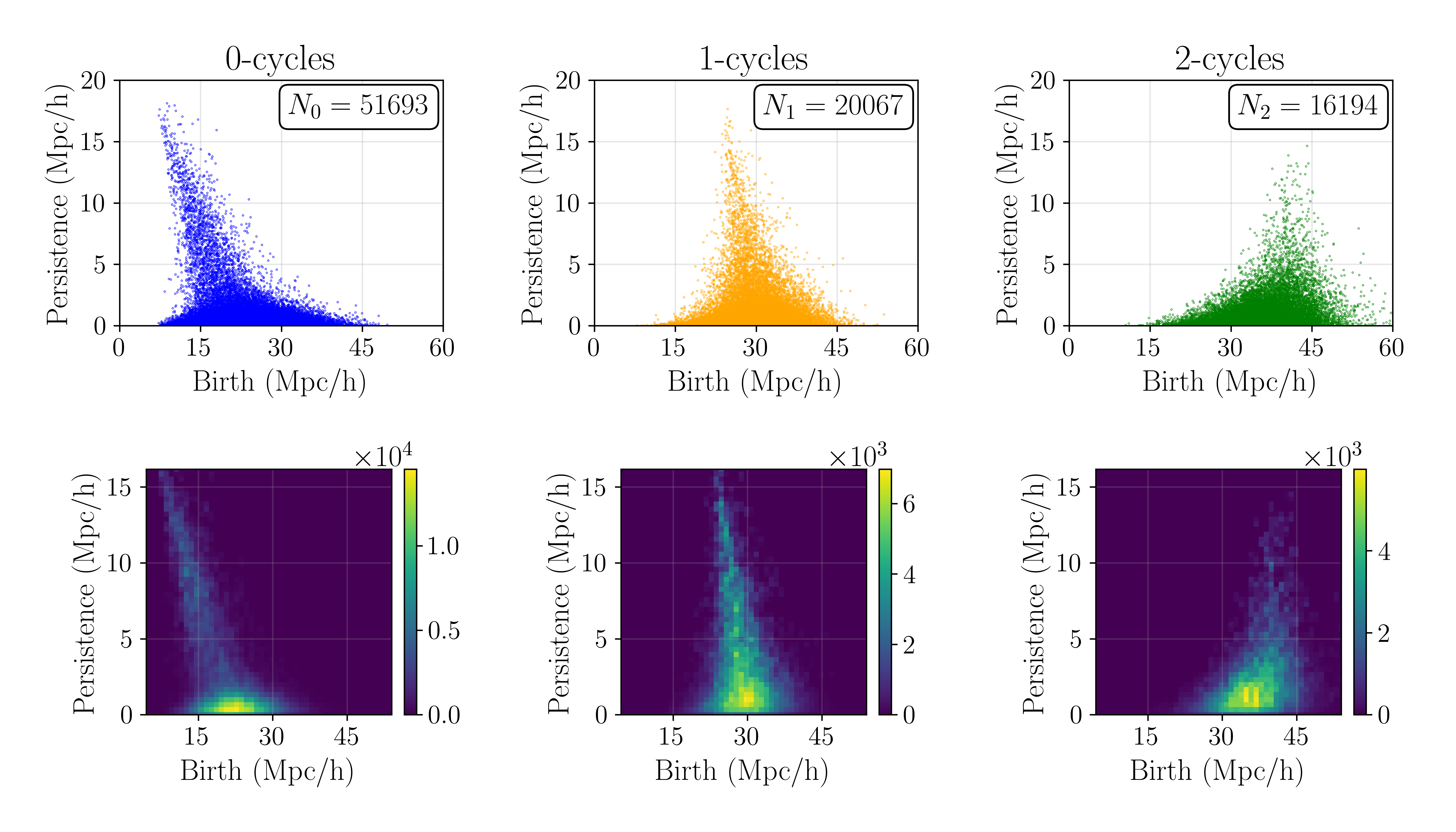}
        \caption{\textbf{Persistence diagrams and images.} \emph{Top Panel:} Persistence diagrams of $0$-, $1$-, and $2$-cycles from the filtration of a \quijote halo catalog with $k=15$ at the fiducial cosmology. $N_p$ is the total number of $p$-cycles in each diagram, i.e., the total number of $p$-cycles once existed in the filtration. \emph{Bottom Panel:} Corresponding persistence images from ``pixelating'' the diagrams.}
        \label{fig:ph}
\end{figure}

When applying persistent homology to $3$-dimensional spatial distributions of dark matter halos, the outputs from a single filtration are three lists of $(\nu_{\rm birth},$ $\nu_{\rm persist})$ pairs,\footnote{We exclude short-lived topological features with $\nu_{\rm persist}\leq0.01\,$Mpc$/h$.} one for each of the three homological dimensions. They can be presented in the form of \emph{persistence diagrams}. In Figure \ref{fig:ph}, we show persistence diagrams representative of the fiducial \quijote halo catalogs along with their corresponding \emph{persistence images}. A persistence image is obtained by first applying a smoothing kernel to every point in a persistence diagram, after which we discretize the birth-persistence plane and sum up all kernel contributions within each pixel. Given that most features are short-lived, we customarily weigh each point by its persistence to emphasize the more persistent features.

\begin{figure}
    \centering
         \includegraphics[width=0.95\textwidth]{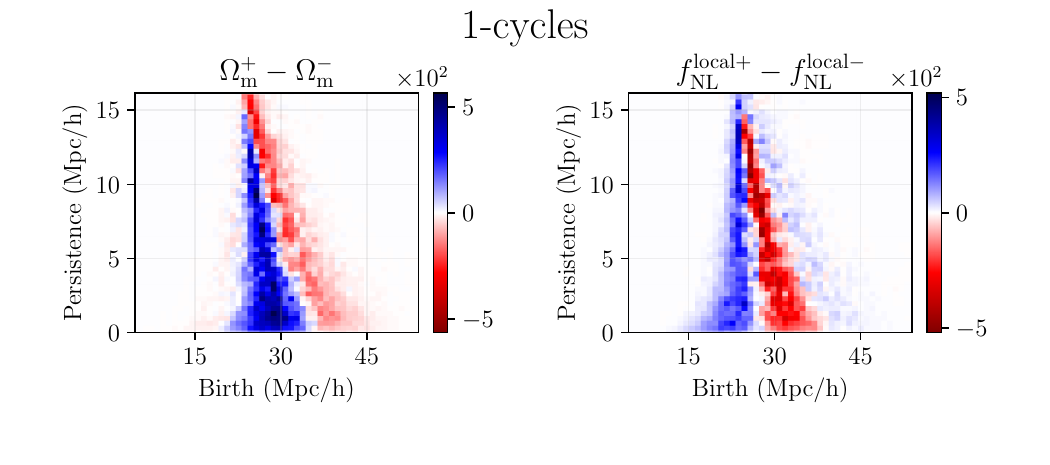}
         \caption{\textbf{Persistence image difference.} To intuitively understand how a cosmological parameter impacts the topology, we can compare two persistence diagrams by computing the persistence image difference. Here, with $k=15$ and using $100$ catalogs, we subtract the mean persistence image of the minus ($\theta^-$) cosmology from that of the plus ($\theta^+$) cosmology for $\Omega_{\rm{m}}$ (left) and $f_{\rm{NL}}^{\rm{local}}$ (right). Regions in blue (red) have more (fewer) $1$-cycles in the $\theta^+$ cosmology. The triangular pattern hints that statistics built from birth and death counts would be informative about the cosmology. Plots for all parameters and homological dimensions are shown in Appendix \ref{app:addplot}.}
        \label{fig:pi}
\end{figure}

Persistence images are useful for intuitively understanding how changes in the point cloud, due to parameters relevant to cosmology in our context, affect the persistence diagrams and, thus, the topology. Specifically, for a parameter $\theta$, we can calculate the difference between the mean persistence images from the ``plus'' ($\theta^+$) and ``minus'' ($\theta^-$) cosmologies. In Figure \ref{fig:pi}, we show two examples: $\Omega_{\rm{m}}$ on the left and $f^{\rm{local}}_{\rm{NL}}$ on the right, both for $1$-cycles. The most distinctive change in the number of features is observed within a range of birth scales. For instance, when $\Omega_{\rm{m}}$ is above (below) its fiducial value, more (fewer) $1$-cycles tend to form at around $27\,$Mpc$/h$ and fewer (more) at scales smaller or larger than that. The opposite behavior is observed for $f^{\rm{local}}_{\rm{NL}}$.\footnote{Plots for all parameters and homological dimensions are shown in Appendix \ref{app:addplot}.} This motivates the construction of a data vector of feature counts in birth (and death) scale bins as our statistic, which we explain in the next section.

\paragraph{Software.}
We use the public code\footnote{\texttt{\url{https://gitlab.com/mbiagetti/persistent\_homology\_lss}}} developed in \cite{Biagetti:2020skr} to compute $\alpha$-DTM$\ell$-filtrations of halo catalogs, as detailed in the previous sections, and \texttt{GUDHI} \cite{gudhi:urm} to perform the persistence calculations.

\section{From Persistence Diagrams to Summary Statistic}\label{sec:summarystat}
A persistence diagram needs to be vectorized to generate a data vector, as it is a $2$D multiset with a variable number of points that are not ordered. The vectorization results in a summary statistic, which is generally a compression of the information originally contained. It is desirable that the compression remains sensitive to cosmological parameter changes, i.e., it retains a substantial amount of cosmological information.

One may flatten the persistence images and concatenate the resulting vectors, but it is unclear if this statistic has a Gaussian likelihood which is required for our Fisher forecast setup\footnote{See \cite{yip:2023} for a less restricted scenario of using persistence images in combination with a convolutional neural network model for cosmological parameter estimation.}. The high dimensionality (number of pixels) of the images also poses difficulties in the estimation of the covariance matrix and the convergence of the forecast constraints due to the limited number of available simulations. Hence, we attempt to devise simpler summary statistics. The triangular pattern of each persistence diagram (Figures \ref{fig:ph} and \ref{fig:pi}) suggests the following statistics recently studied in the literature:
\begin{itemize}
    \item \textit{$1$-dimensional curves:} In \cite{Biagetti:2020skr}, we scanned each persistence diagram in different directions, thereby defining empirical distribution functions of $\nu$. For example, along the birth axis, the function is the count of all features born at scales larger than $\nu$.
    \item \textit{Histograms of births and deaths:} In \cite{Biagetti:2022qjl}, we used histograms of $\nu_{\rm birth}$ and $\nu_{\rm death}$ values. These are summaries simpler than the $1$-dimensional curves since counts from different scales are not stacked, and their responses to changes in the cosmology typically converge better.
\end{itemize}

In other contexts, there exist several proposals for this map \cite{carriere2015stable,bubenik2015statistical,adams2017persistence,kalivsnik2019tropical}. A promising approach would be to parameterize the map by a neural network, imposing that the network is permutation invariant to the input topological features \cite{zaheer2017deep,carriere2020perslay}. While these techniques require detailed investigations of their applications to real datasets, in this paper, we opt for the simpler approach of constructing histograms, being aware that this method of compression is likely suboptimal.

\subsection{Description of the data vector}\label{sec:datavector}
We build our histogram data vector by counting the number of features in birth and death bins. From a single filtration, a set of $6$ histograms\footnote{Note that the minimum and maximum scales probed by these histograms are fixed regardless of the cosmology, and we remove values above the $99.9$th percentile (also death values below the $0.01$th percentile) to avoid effects due to sparsity.} can be obtained: $3$ from the birth distributions of the $0$-, $1$-, and $2$-cycles ($B_0$, $B_1$, $B_2$), and the other $3$ from the death distributions ($D_0$, $D_1$, $D_2$). As elaborated in Section \ref{sec:cgvf}, we vary the nearest-neighbor parameter $k$ (Eq. \eqref{eq:dtm}) and use the joint vector of $6$ sets of these histograms, corresponding to $6$ values of $k\in\{1,5,15,30,60,100\}$. Although this construction of the data vector is intentionally kept generic and not optimized for the tightest possible constraints in this work, we can show that each component contributes a fair amount of independent information, see Appendix \ref{app:howinform}.

\begin{figure}
    \centering
        \includegraphics[trim={0cm 0cm 0cm 0cm},width=1\textwidth, clip]{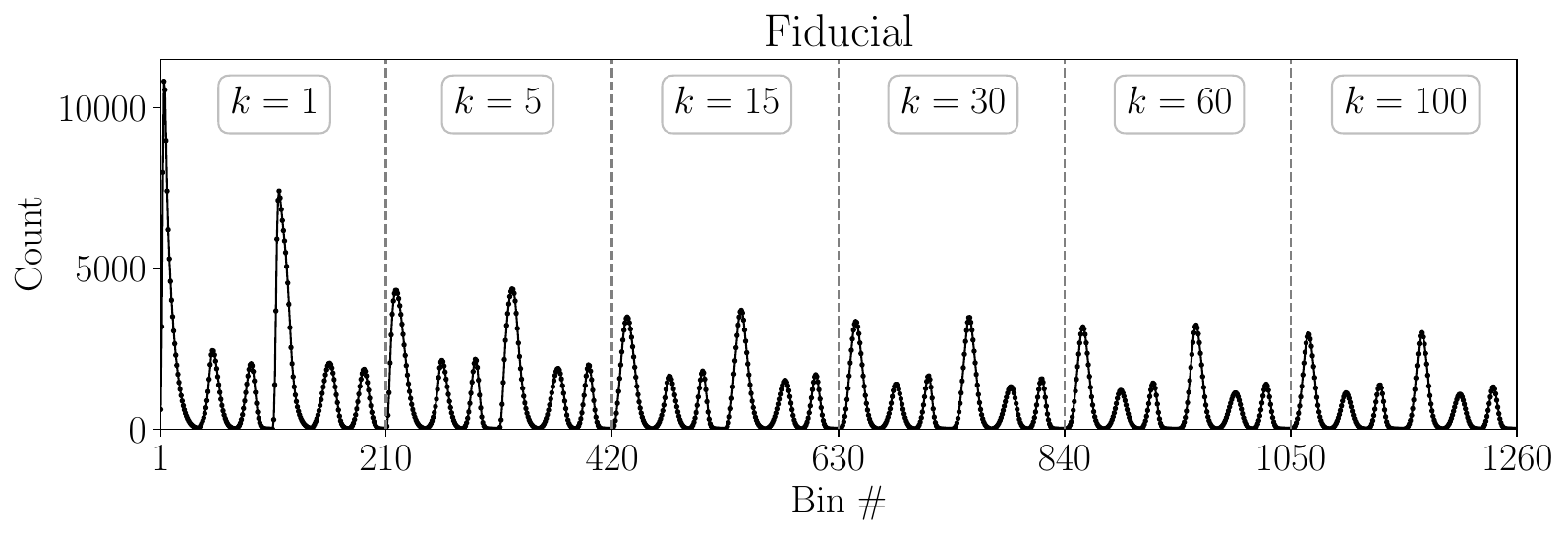}
        \caption{{\bfseries Full data vector.} Displayed here for the fiducial cosmology is our full data vector, which is the concatenation of $6$ sets of histograms with the nearest-neighbor parameter $k\in\{1,5,15,30,60,100\}$. For each value of $k$, that subset of the data vector contains a total of $6$ histograms: one from the birth distribution and another from death for each of the $3$ homological dimensions.}
        \label{fig:fulldv}
\end{figure}

\begin{figure}
    \centering
        \includegraphics[trim={0.5cm 0cm 0.5cm 0cm},width=1\textwidth, clip]{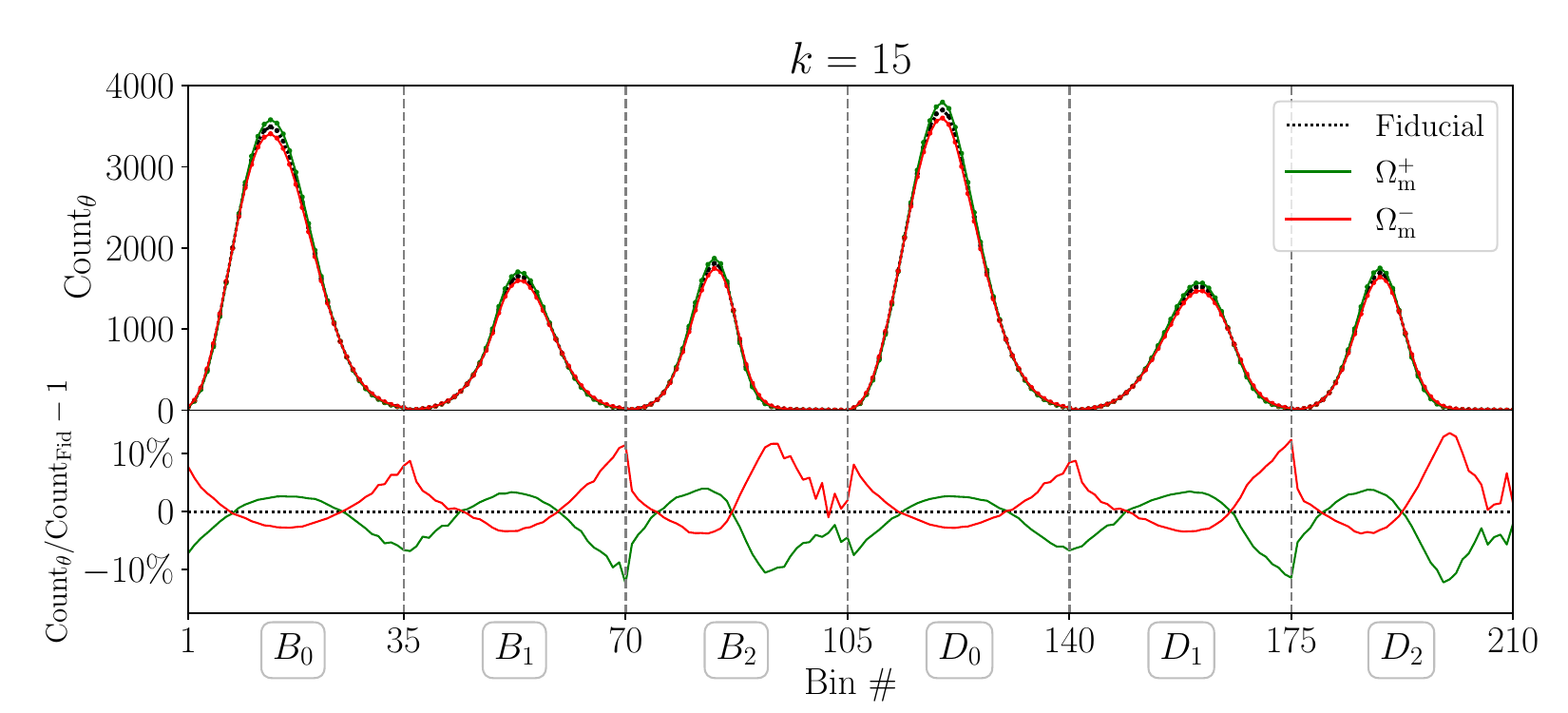}
        \caption{{\bfseries Varying cosmologies.} \emph{Top Panel:} Subset of the full data vector corresponding to $k=15$ for the fiducial (dotted black), $\Omega^{+}_{\rm{m}}$ (solid green), and $\Omega^{-}_{\rm{m}}$ (solid red) cosmologies, averaged over $14500$ (fiducial) and $1500$ ($\Omega^{\pm}_m$) realizations. Vertical dashed lines separate histograms of the birth ($B_p$) and death ($D_p$) values for $p$-dimensional cycles. \emph{Bottom Panel:} Percentage deviations of the $\Omega^{\pm}_{\rm{m}}$ cosmologies from the fiducial one.}
        \label{fig:datavector}
\end{figure}

For visualization, we show
\begin{itemize}
    \item The full data vector for the fiducial cosmology (Figure \ref{fig:fulldv}), which is the concatenation of the $6\times6=36$ histograms, averaged over $14500$ realizations.
    \item The subset of the data vector corresponding to $k=15$ for the fiducial and varying $\Omega_{\rm{m}}$ cosmologies (Figure \ref{fig:datavector}), averaged  over $14500$ (fiducial) and $1500$ ($\Omega^{\pm}_{\rm{m}}$) realizations. As shown in the figure, within each $k$, the histograms of birth and death values are concatenated in the order of $\{B_0, B_1, B_2, D_0, D_1, D_2\}$. The lower panel shows the percentage deviations of the $\Omega^{\pm}_{\rm{m}}$ cosmologies from the fiducial one.
\end{itemize}
We have tested various data vector sizes, ranging from $360$ to $11160$ for the total number of bins $N_{\rm bin}$. We choose $N_{\rm bin}=1260$ for our reported results, a decision driven by balancing the Gaussianity of our statistic, the convergence of the numerical derivatives, and the need for powerful constraints. We expand on these considerations in Appendix \ref{app:tests}.

\begin{figure}
    \centering
        \includegraphics[trim={0.5cm 0cm 0.5cm 0cm},width=1\textwidth, clip]{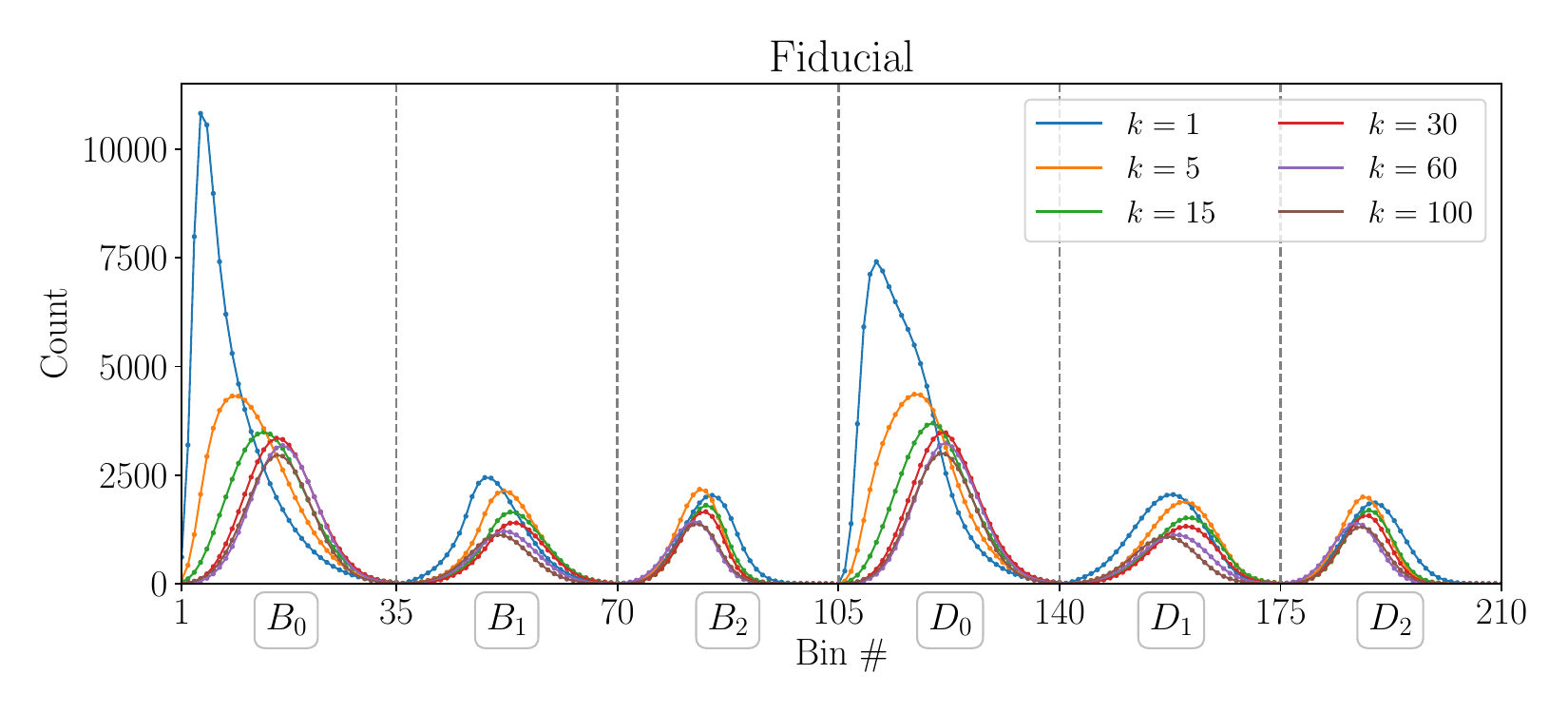}
        \caption{{\bfseries Varying $k$ values.} Displayed here are subsets of the fiducial full data vector corresponding to $k=1$, $5$, $15$, $30$, $60$, and $100$. The skewness of the histograms varies and the peaks become lower as $k$ increases.}
        \label{fig:kdv}
\end{figure}

\paragraph{Impact of $k$.}
In Figure \ref{fig:kdv} we compare the $6$ sets of histograms from filtrations with the different values of the nearest-neighbor parameter $k$. As $k$ increases, the skewness of the histograms varies, and the peaks become lower in accordance with Figure \ref{fig:kEffect} that a larger $k$ value causes persistent homology to track fewer but larger-scale features. Note that the histograms do not share the same bins across $k$ values, otherwise we would also see the peaks shift towards the larger-scale bins (to the right) as $k$ increases. The $6$ values of $k$ are chosen empirically, while in theory the information maximizing combination of values can be found from optimizing the forecast result, which is addressed in the related work \cite{Karthik}.

\subsection{Covariance}
\begin{figure}
    \centering
        \includegraphics[trim={1.5cm 0cm 0cm 2.5cm}, width=1\textwidth, clip]{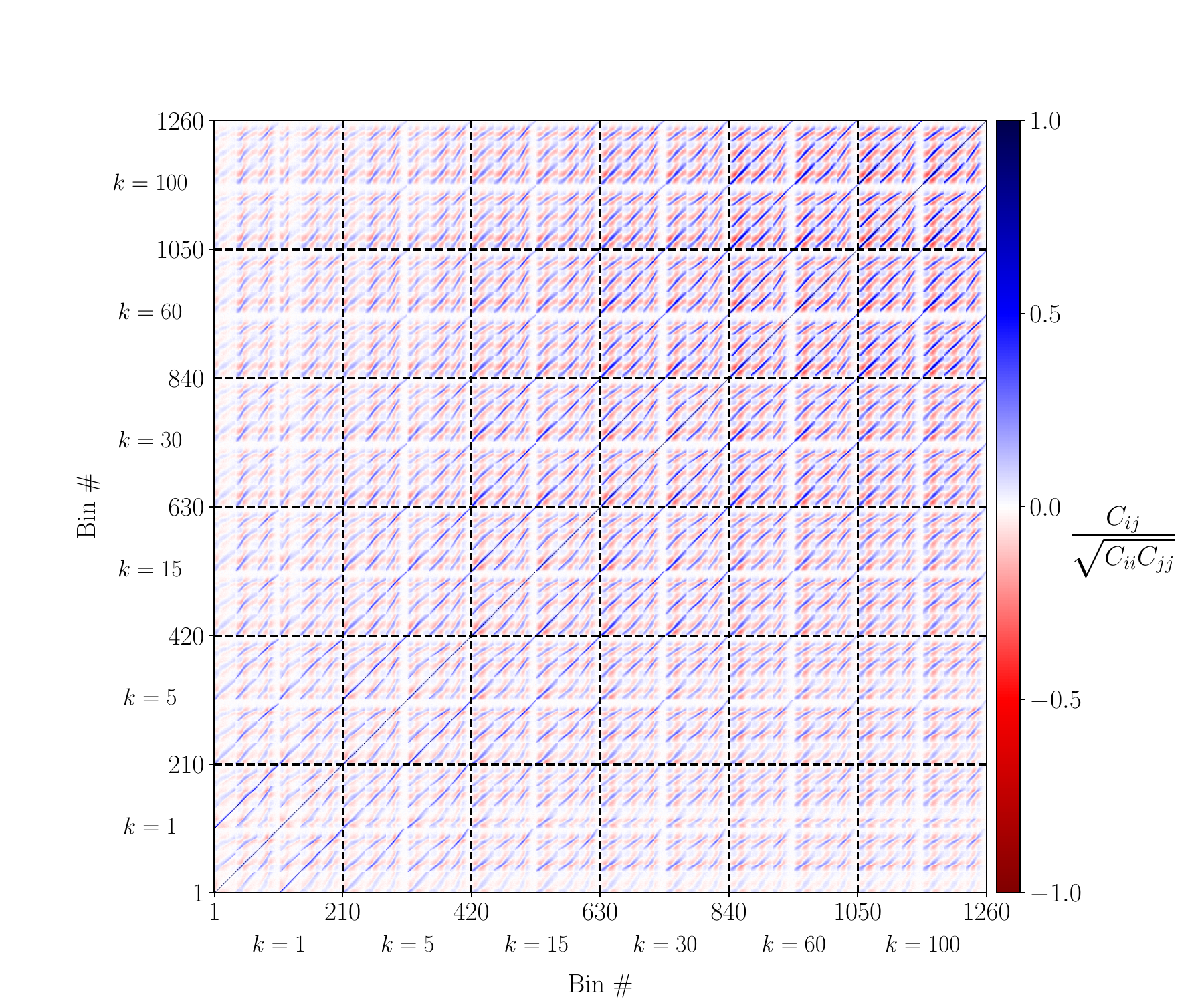}
        \caption{{\bfseries Correlation matrix of the full data vector at the fiducial cosmology.} Positive (negative) correlations are in blue (red). Each block, outlined by dashed lines, contains the correlations between two subsets of the data vector corresponding to the same or different $k$ values. Larger-$k$ bins are more strongly correlated, indicating that the amount of information extractable is saturating.}
        \label{fig:corrmatrix}
\end{figure}

In Figure \ref{fig:corrmatrix} we show the correlation matrix of our full data vector at the fiducial cosmology, defined as
\begin{equation}
    r_{ij} = \frac{C_{ij}}{\sqrt{C_{ii}C_{jj}}},
\end{equation}
where $C_{ij}$ is the covariance matrix
\begin{equation}
    C_{ij} = \frac{1}{N_{\rm sim}-1} \,\sum_{N=1}^{N_{\rm sim}} [(D_{\rm Fid})_i - \langle D_{\rm Fid} \rangle_i][(D_{\rm Fid})_j - \langle D_{\rm Fid} \rangle_j],
\end{equation}
with $N_{\rm sim} = 14500$ fiducial realizations and $\langle D_{\rm Fid} \rangle$ being the mean data vector over those realizations. The diagonal stripe pattern in the figure is due to the fact that the $36$ peaks (one from each histogram) in the data vector shift together across bins. For example, delays in the births of $0$-cycles in general cause delays in their deaths, as well as in the births and deaths of higher-dimensional features. We also observe that the larger $k$ bins are more strongly correlated, indicating that the amount of information extractable by including more larger values of $k$ is getting saturated.

\section{From Summary Statistic to Cosmological Constraints}\label{sec:results}
The \quijote suite is designed for numerically evaluating the information content of a summary statistic via the Fisher information matrix at the fiducial cosmology. If the likelihood function of a data vector $D$ is Gaussian, the Fisher matrix elements are\footnote{Note that we conservatively neglect the dependence of the covariance matrix on the parameters, see footnote $9$ of \cite{Biagetti:2022qjl} for a detailed explanation.} 
\begin{equation}\label{eq:Fisher}
    F_{ij}=D_{,\theta_i}^T C^{-1}D_{,\theta_j}
\end{equation}
where $C^{-1}$ is the inverse of the covariance matrix, and $D_{,\theta_i}$ denotes the derivative of $D$ with respect to the parameter $\theta_i$. $C^{-1}$ and $D_{,\theta_i}$ are computed numerically using the fiducial and varying cosmologies of \quijote, respectively.\footnote{We include the Hartlap factor \cite{Hartlap:2006kj} for the unbiased estimation of $C^{-1}$, i.e., $C^{-1}_{\rm unbiased}=\frac{N_{\rm sim}-N_{\rm bin}-2}{N_{\rm sim}-1} C^{-1}$. For the numerical estimation of $D_{,\theta_i}$, see Appendix \ref{app:nuderi}.} Therefore, this symmetric matrix is of the size $7\times 7$ for our $7$ cosmological parameters $\theta_{\rm cosmo} = \{ \Omega_{\rm{m}}, \Omega_{\rm{b}}, h, n_{\rm{s}}, \sigma_8, \sum m_\nu,w\}$. When considering primordial non-Gaussianity, the size is $10\times 10$, since we include all three shapes at the same time and also marginalize over the cosmological parameters. The diagonal of the inverse of the Fisher matrix contains (the lower bound on) the variance of, hence the marginalized constraints on, the parameters.

The focus of this work is to compare persistent homology against the conventional low-order correlation functions, i.e., the $2$- and $3$-point statistics. Specifically, the two data vectors are, respectively, from our persistent homology pipeline (hereafter labeled as $PH$) and from combining, up to $k = 0.3\,h/$Mpc, the power spectrum monopole and quadrupole as well as the bispectrum monopole (labeled as $P_0+P_2+B_0$). We run consistency checks on our results and various components of the calculations, which we present in Appendix \ref{app:tests}.

\subsection{Constraints on cosmological parameters}
\begin{figure}
    \centering
        \includegraphics[width=1.05\textwidth]{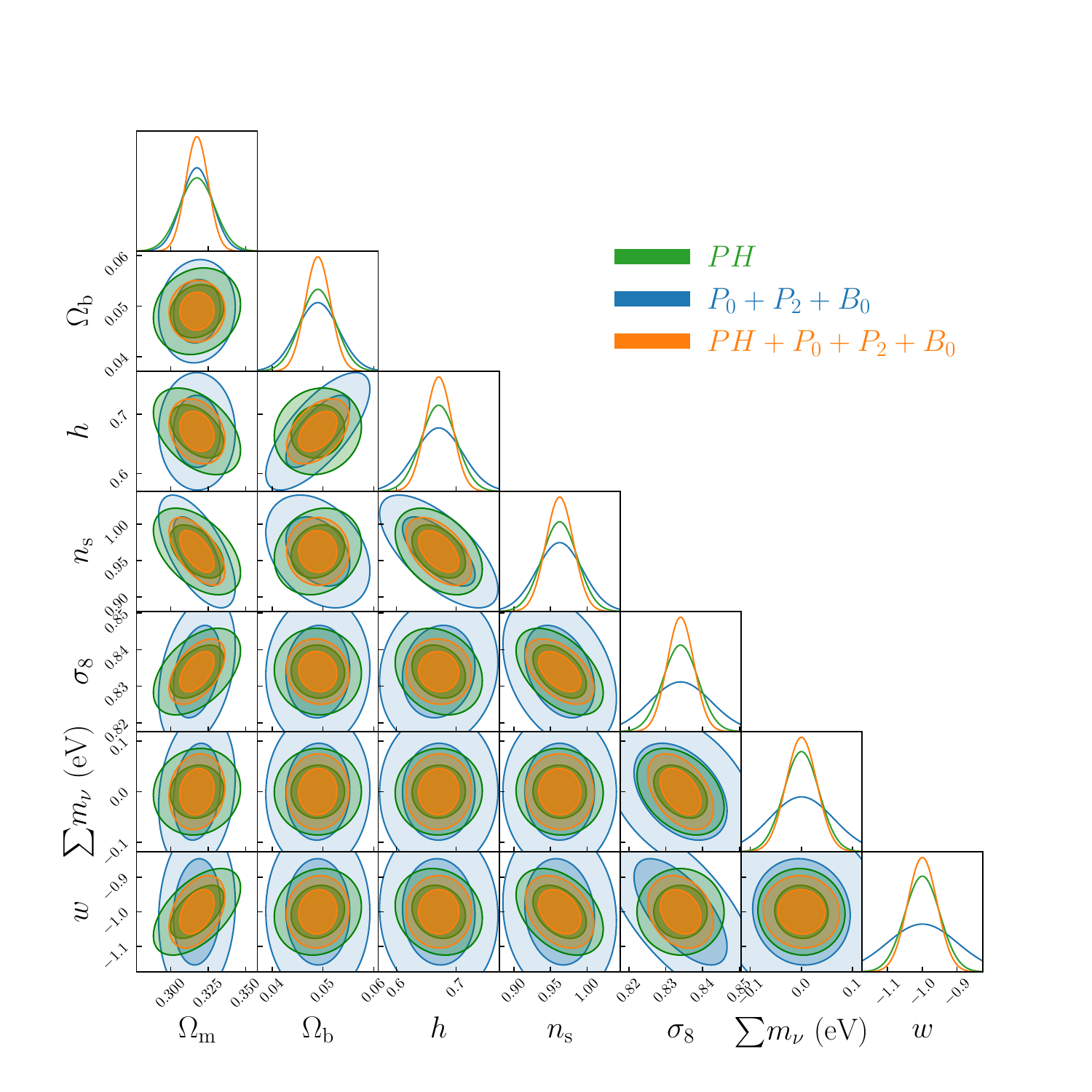}
        \caption{\textbf{Confidence contours for cosmological parameters.} A comparison of the marginalized contours marking the $68\%$ and $95\%$ confidence regions from persistent homology (\textcolor{ForestGreen}{$PH$}), redshift space halo power spectrum monopole and quadrupole and bispectrum monopole (\textcolor{RoyalBlue}{$P_0+P_2+B_0$}), and combining the two (\textcolor{YellowOrange}{$PH+P_0+ P_2+B_0$}). The degeneracy directions differ for the two statistics, and as a result contours from combining are considerably more constraining for most of the parameters.}
        \label{fig:combined}
\end{figure}

\begin{figure}
    \centering
    \includegraphics[width=0.8\textwidth]{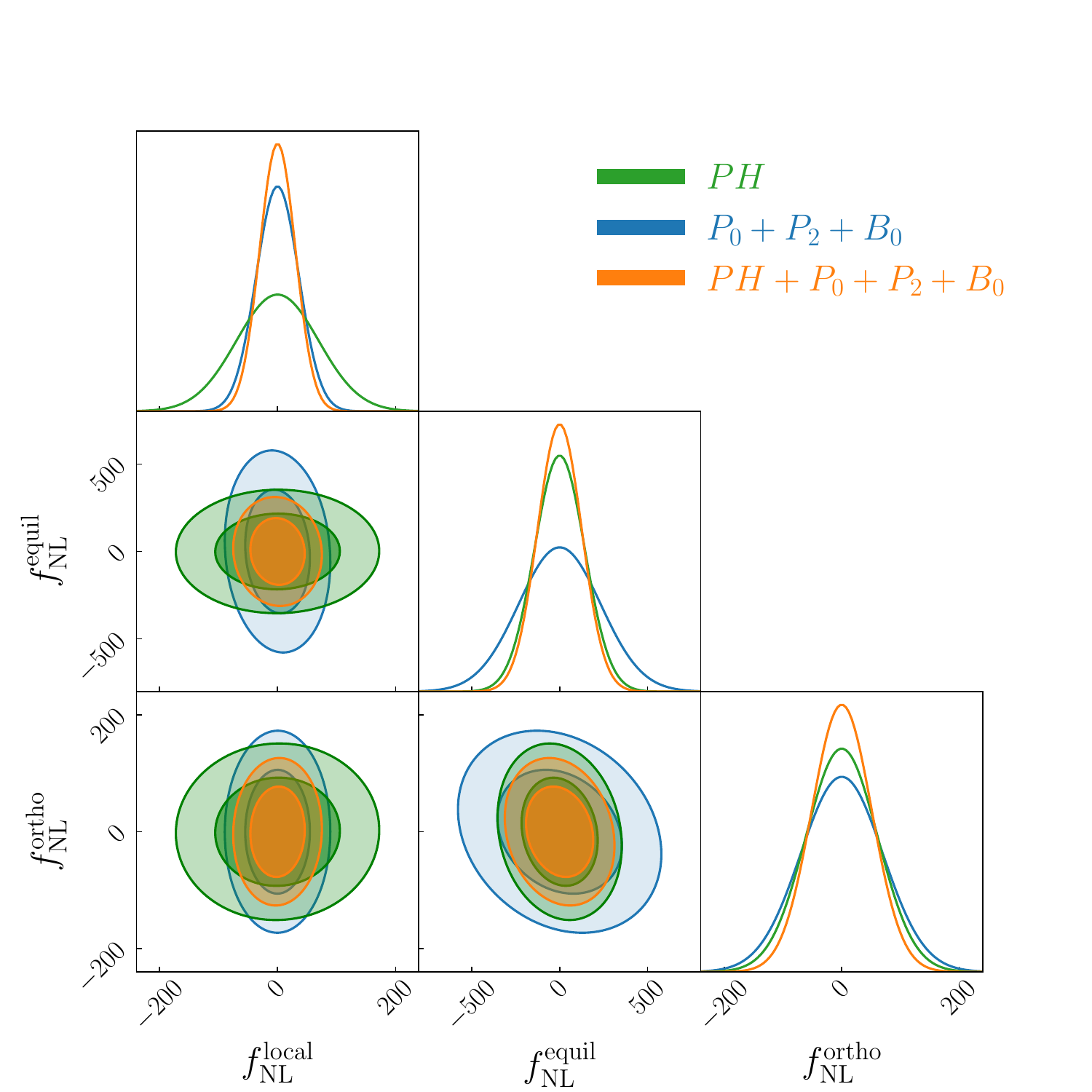}
    \caption{\textbf{Confidence contours for primordial non-Gaussianity amplitudes.} A comparison of the marginalized contours marking the $68\%$ and $95\%$ confidence regions from persistent homology (\textcolor{ForestGreen}{$PH$}), redshift space halo power spectrum monopole and quadrupole and bispectrum monopole (\textcolor{RoyalBlue}{$P_0+P_2+B_0$}), and combining the two (\textcolor{YellowOrange}{$PH+P_0+ P_2+B_0$}).}
    \label{fig:png}
\end{figure}

We show the confidence contours for the cosmological parameters in Figure \ref{fig:combined}.\footnote{The attentive reader may have noticed that we are allowing for unphysical negative values of the total neutrino mass. To obtain physically sound posteriors from sampling, one should discard samples that are associated with negative total neutrino masses. This implies that the resulting posteriors will all be non-Gaussian, strictly speaking. In practice, we have verified that this procedure modifies the contours for other parameters only negligibly. Therefore, we stick to symmetrical distributions of total neutrino mass values for illustrative purposes.} Note that each contour is marginalized over all other cosmological parameters, and the three primordial non-Gaussianity amplitudes are fixed at their fiducial value $0$. The $1$-$\sigma$ constraints from persistent homology are tighter than those from the joint power spectrum and bispectrum statistic by $23-50\%$ for all cosmological parameters except $\Omega_{\rm{m}}$.\footnote{The percentage improvement is defined as $(\sigma_{P_0+P_2+B_0}-\sigma_{PH})/\sigma_{P_0+P_2+B_0}$.} We claim that this is primarily due to persistent homology assessing information of high-order correlators, as topological features are built from associating multiple points.

While it is unclear how we can assign to our summary statistic the precise range of scales it probes, we propose that most of the information comes from scales above $2\pi/(0.3\,h/$Mpc$)=21\,$Mpc$/h$, which is the minimum scale for the joint power spectrum and bispectrum statistic. This is evident when examining the persistence images in Figure \ref{fig:ph}, where most $1$- and $2$-cycles are born between $\nu_{\rm birth}\approx25-40\,$Mpc$/h$. As detailed in Section \ref{sec:cgvf}, $\nu_{\rm birth}$ corresponds to half of the length of the $1$-simplex\footnote{The filtration parameter $\nu$ is roughly the radius of the growing ball around each vertex, and a $1$-simplex is added to the simplicial complex when two balls touch.} that triggers the creation of a cycle. The length scale of the created cycle is expected to be larger than that of the $1$-simplex; therefore, the scales of the $1$- and $2$-cycles from which we extract information exceed $2\nu_{\rm birth}\approx50-80\,$Mpc$/h$.\footnote{Strictly speaking, since $\nu_{\rm birth}$ values in an $\alpha$-DTM$\ell$-filtration are delayed by the DTM function (Eq. \ref{rx}), the range of minimum scales quoted here is an overestimation. Nevertheless, as stated the length scales of the cycles are expected to be larger than $2\nu_{\rm birth}$, and there is also a considerable margin between $21$ and $50\,$Mpc$/h$.} For the $0$-cycles, $\nu_{\rm birth}$ instead corresponds inversely to the number density around a halo. Given that from Figure \ref{fig:ph} most $0$-cycles are born between $\nu_{\rm birth}\approx20-25\,$Mpc$/h$, they are consistent with information from scales $\gtrsim21\,$Mpc$/h$.\footnote{The analysis here focuses on values from Figure \ref{fig:ph} where the nearest-neighbor parameter is $k=15$. Yet the argument applies to the other $k$ values, except for the $0$-cycles from $k=1$ and $5$, which indeed have small $\nu_{\rm birth}$ values. However, there is no cause for alarm, as they contribute only limitedly to our data vector (see Appendix \ref{app:howinform}).\label{fn:noalarm}}.

Another observation is that the parameter degeneracies for our statistic are in directions fairly different from those for the joint power spectrum and bispectrum statistic. This implies that we can break degeneracies by combining the two statistics and obtain constraints that are tightened by $12-33\%$. We show the contours from combining in orange in the same Figure and summarize all $1$-$\sigma$ constraints in Table \ref{tab:results}. Here we point out a few major observations:

\begin{itemize}
    \item $PH$ shows stronger degeneracies in the $\sigma_8$-$\Omega_{\rm m}$, $h$-$\Omega_{\rm m}$, and $w$-$\Omega_{\rm m}$ planes with respect to $P_0+ P_2+B_0$. This might be due to the fact that persistent homology is more sensitive to the local number density of points (or more specifically the inter-halo distance), which changes in a degenerate way for these parameters.
    \item $PH$ shows weaker degeneracies in the $\sigma_8$-$w$, $\Omega_{\rm b}$-$h$, and $\Omega_{\rm b}$-$n_{\rm s}$ planes. They allow for the gains in constraining power from combining the statistics and deserve future investigations.
    \item Overall, constraints on $\sigma_8$, $w$ and $\sum m_\nu$ are considerably improved with respect to $P_0+ P_2+B_0$. We reiterate that given the simplistic setup, these constraints should not be taken at face value, but only in the relative sense. Nevertheless, especially in the case of massive neutrinos, such an improvement may be expected given the higher sensitivity of our statistic to mildly non-linear scales, as noted above.
\end{itemize}

\subsection{Constraints on primordial non-Gaussianity}
We show the confidence contours for the primordial non-Gaussianity amplitudes in Figure \ref{fig:png}. Each contour is marginalized over all cosmological parameters and the one primordial non-Gaussianity amplitude remaining. While the volume of the \quijote catalogs is insufficient to give interesting constraints for these non-Gaussianity shapes, our analysis sheds light on whether a summary statistic that contains information of high-order correlators helps for the equilateral and orthogonal shapes, which do not have a strong scale dependence at large scales unlike the local one (see \cite{Biagetti:2019bnp} for a recent review on the effect of primordial non-Gaussianity on biased tracers). We observe in Figure \ref{fig:png} that constraints from persistent homology are indeed tighter for $f_{\text{NL}}^{\text{equil}}$ and $f_{\text{NL}}^{\text{ortho}}$. On the other hand, $f_{\text{NL}}^{\text{local}}$ is as expected best constrained by the joint power spectrum and bispectrum statistic, since most information is in the large scales, while there are only so many large-scale holes for persistent homology to track. As for breaking parameter degeneracies, combining the two statistics tightens the constraints by $11-17\%$, which are more modest improvements in comparison to those for the cosmological parameters.

\begin{table}[t]
\begin{centering}
\begin{tabular}{ |P{2cm}|P{1cm}|P{3.7cm}|P{3.7cm}|P{3.7cm}| } 
    \hline 
        $\theta$ & \tiny{Fiducial} & \textcolor{ForestGreen}{$PH$} & \textcolor{RoyalBlue}{$P_{0}+P_{2}+B_{0}$} & \textcolor{YellowOrange}{$PH+P_{0}+P_{2}+B_{0}$}\\ 
    \hline
    \hline
        $\Omega_{\text{m}}$ & $0.3175$ & $\pm0.012$ & $\pm0.010$ & $\pm0.008$\\
        $\Omega_{\text{b}}$ & $0.049$ & $\pm0.003$ & $\pm0.004$ & $\pm0.002$\\
        $h$ & $0.6711$ & $\pm0.029$ & $\pm0.040$ & $\pm0.022$\\
        $n_{\text{s}}$ & $0.9624$ & $\pm0.024$ & $\pm0.031$ & $\pm0.019$\\
        $\sigma_8$ & $0.834$ & $\pm0.005$ & $\pm0.008$ & $\pm0.004$\\
        $\sum m_\nu$ (eV) & $0$ & $\pm0.034$ & $\pm0.063$ & $\pm0.030$\\
        $w$ & $-1$& $\pm0.051$ & $\pm0.101$ & $\pm0.042$\\
    \hline
    \hline
        $f_{\text{NL}}^{\text{local}}$ & $0$& $\pm69$ & $\pm36$ & $\pm30$\\
        $f_{\text{NL}}^{\text{equil}}$ & $0$& $\pm142$ & $\pm233$ & $\pm126$\\
        $f_{\text{NL}}^{\text{ortho}}$ & $0$& $\pm61$ & $\pm70$ & $\pm51$\\
    \hline
\end{tabular}
\par\end{centering}
\caption{\label{tab:results}\textbf{$1$-$\sigma$ constraints.} $1$-$\sigma$ Fisher constraints from persistent homology (\textcolor{ForestGreen}{$PH$}), redshift space halo power spectrum monopole and quadrupole and bispectrum monopole (\textcolor{RoyalBlue}{$P_0+P_2+B_0$}), and combining the two (\textcolor{YellowOrange}{$PH+P_0+ P_2+B_0$}). These results are for a $1\,($Gpc$/h)^3$ volume at $z=0.5$ for tracers with a number density of $\bar{n}\approx9.88\times10^{-5}\,(h/{\rm Mpc})^3$. The cosmological parameter constraints are marginalized over all parameters except the three primordial non-Gaussianity amplitudes, which are set to $0$. Constraints on primordial non-Gaussianity amplitudes are marginalized over all parameters.}
\end{table}

\section{Conclusions and Future Directions}\label{sec:conclusions}
In a series of recent papers \cite{Biagetti:2020skr,Biagetti:2022qjl,yip:2023}, we have explored a class of summary statistics based on persistent homology, a formalism within the field of computational topology, which we customize for quantifying the morphology of the cosmic web as a distribution of clusters, loops, and voids across scales. In this work, with a histogram-based statistic, we have measured its information content through a Fisher analysis and reported constraints on cosmological parameters and primordial non-Gaussianity amplitudes.

We have presented a comparison between these constraints from persistent homology and those from statistics currently used in cosmological survey analyses, namely the power spectrum and the bispectrum. We have found that persistent homology outperforms the conventional statistics for 8 out of the 10 parameters by margins of $13-50\%$. Combining the methods breaks degeneracies between parameters and further tightens the constraints. We claim that this is expected, since persistent homology probes high-order correlations complementary to the $2$- and $3$-point statistics through topological features constructed from many vertices. Our results strongly motivate further investigations of utilizing persistent homology in constraining parameters with actual survey data.

In this perspective, this work suggests several directions for future work:

\begin{itemize}
    \item {\bfseries Maximizing information.} Data compression with persistent homology is subject to various choices leading to the final summary statistic. There is the freedom to choose a specific filtration, set the hyperparameters within the filtration, and then define a vectorization strategy. Many of these are ``discrete'' choices, i.e., there are finitely many options in, e.g., choosing a particular filtration or vectorization. There are also hyperparameters to fine tune, such as the nearest-neighbor parameter $k$ relevant to this work. To determine the best choices and values, we should optimize the analysis pipeline for maximized information, by defining a well-behaved figure of merit that measures the information content, and an algorithm to update the pipeline towards the optimal compression. Then it seems natural to use (the negative of) the Fisher information as the loss function of a neural network that parameterizes the pipeline, as proposed in the recent work \cite{charnock2018}. We are currently implementing similar methods to optimize the mixing parameters $q$ and $p$ in the $\alpha$-DTM$\ell$-filtration \cite{Karthik}.
    \item {\bfseries Inference pipeline.} The Fisher constraints obtained in this work are only parameter uncertainty estimates. When analyzing real data, we would want to infer parameter values from our summary statistic measured on the dataset. As a recent example, \cite{yip:2023} implemented a pipeline utilizing a convolutional neural network model to map persistence images to $\sigma_8$ and $\Omega_{\rm m}$, and the result is encouraging and in agreement with ours. However, to include many more parameters for a full-fledged pipeline appears computationally infeasible, as it would require an enormous number of simulations, either for the training dataset or for numerically evaluating the likelihood function in a more standard setup. While this has always been a problem for any simulation-based inference approaches like ours, we suggest mitigating it by employing persistent homology for data compression. For instance, it may be as reasonably effective but overall much less costly, to first calculate persistence diagrams and then to use them as inputs for a much smaller machine learning model, than to have a huge model trained directly on catalogs each containing hundreds of thousands of halo positions. Additionally, various stability theorems state that persistent homology outputs are stable against small perturbations in the input data \cite{cohen2007stability}, which may justify the use of fast simulations to drastically reduce computational time. To summarize, we propose incorporating persistent homology into inference pipelines for increased efficiency.
    \item {\bfseries Realistic galaxy survey data.} The construction of our statistic from persistent homology is well-defined and can be readily applied to any cosmological point cloud simulated in a box, such as a galaxy catalog at a particular redshift. However, in reality, galaxies are observed in lightcones and each survey is characterized by its geometry and systematics. As an example, analyses of the BOSS survey data are plagued by so-called fiber collisions, which affect the determination of redshifts and, consequently, the modeling of the galaxy redshift space power spectrum \cite{Hahn:2016kiy}. On the other hand, persistent homology might be insensitive to certain systematics; we expect those that merely cause slight displacements of galaxies at the map level will not significantly affect our statistic, thanks to the stability theorems mentioned earlier. However, other ones that lead to systematic removal or addition of vertices could have substantial, unmitigable impacts. In general, a thorough survey-specific study of the systematics is necessary when working with real data. Moreover, it might be beneficial to characterize in detail our framework as a function of galaxy properties, such as luminosity, color, and mass, and determine the effects of assembly bias on our statistic.
\end{itemize}

As a final comment, we must be aware that our method suffers the same challenge of robustness that all simulation-based methods do: the information we extract from a model is only as reliable as the simulations themselves are, and it is difficult to distinguish cosmological from non-cosmological information at scales close to the accuracy scale of the simulations. This problem is likely worsened when we include baryons. A recent example in \cite{Villaescusa-Navarro:2021pkb}, where convolutional neural networks are trained on gas temperature maps for cosmological parameter inference, found that a network trained on one simulation, the IllustrisTNG \cite{Pillepich:2017jle}, fails dramatically when applied to maps generated from another simulation suite, the \textsc{simba} simulations \cite{Dave:2019yyq}.

\section*{Acknowledgements}
The authors thank Emilio Bellini and Juan Calles for their help in testing the galaxy mock catalogs, which did not make the final cut of this project as explained in Appendix \ref{app:sancho}. The authors thank William Coulton for discussions on the convergence of Fisher forecasts. The work of J.H.T.Y. and G.S. is supported by the U.S. Department of Energy, Office of Science, Office of High Energy Physics under Award Numbers DE-SC-0023719 and DE-SC-0017647. M.B. is supported by the Programma Nazionale della Ricerca (PNR) grant J95F21002830001 with title ``FAIR-by-design".

Numerical computations were performed 1) under the NWO project ``Topological echoes of primordial physics in cosmological observables.'' on the Dutch National Computing facilities, and 2) using the computing resources and assistance of the University of Wisconsin-Madison Center for High Throughput Computing (CHTC) \cite{https://doi.org/10.21231/gnt1-hw21} in the Department of Computer Sciences. The CHTC is supported by UW-Madison, the Advanced Computing Initiative, the Wisconsin Alumni Research Foundation, the Wisconsin Institutes for Discovery, and the National Science Foundation, and is an active member of the OSG Consortium, which is supported by the National Science Foundation and the U.S. Department of Energy's Office of Science.

\appendix
\section{Numerical Derivatives}\label{app:nuderi}
\begin{figure}
    \centering
        \includegraphics[trim={0.5cm 0cm 0.5cm 0cm},width=1\textwidth, clip]{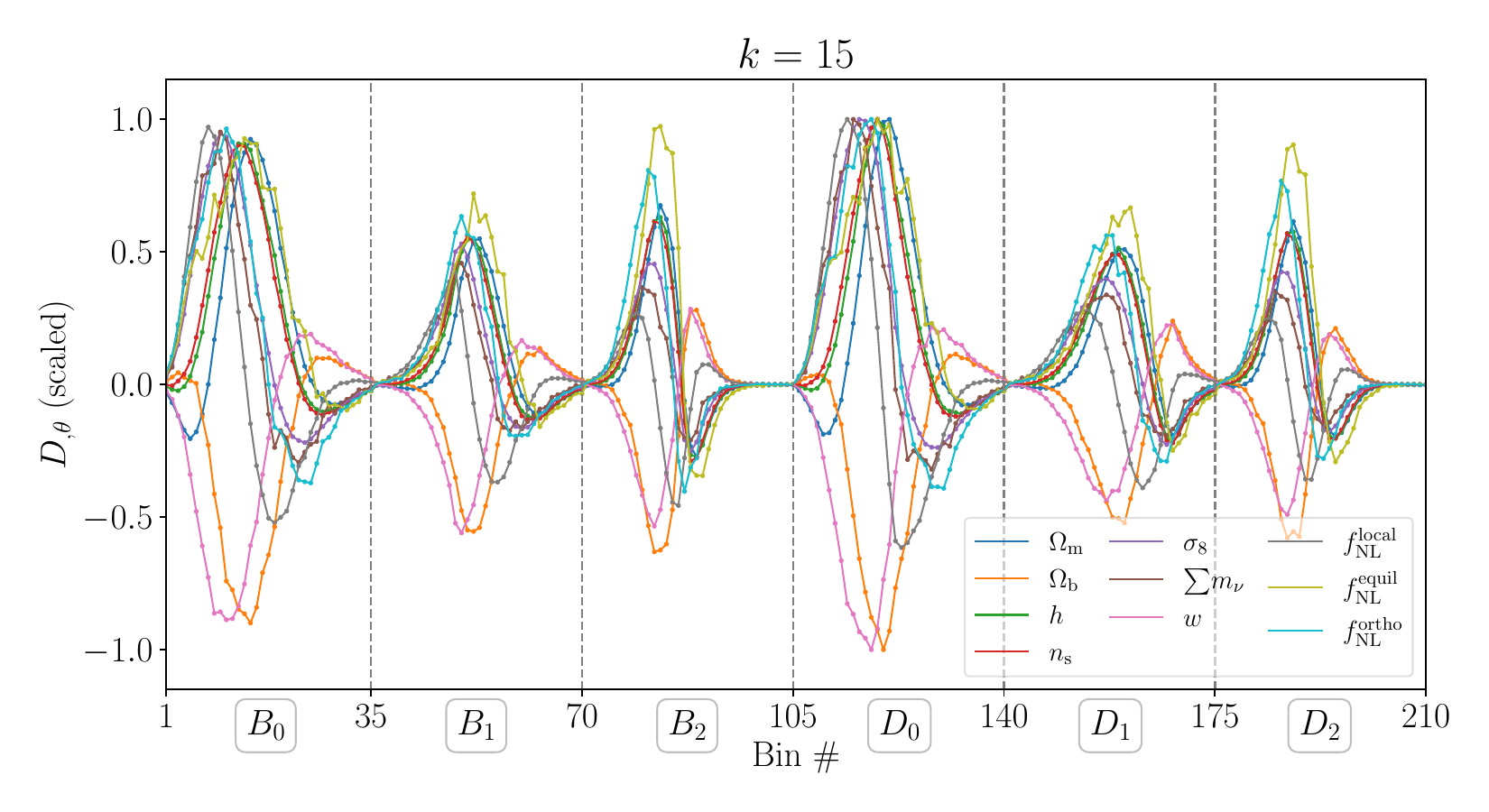}
        \caption{{\bfseries Numerical derivatives of the data vector.} The numerical derivatives ($k=15$) displayed here are scaled between $[-1, 1]$ for illustrative purposes.}
        \label{fig:deri}
\end{figure}

The data vector derivatives required for the Fisher matrix (Eq. \eqref{eq:Fisher}) are computed via the finite difference scheme:
\begin{equation}\label{eq:nuderi1}
    D_{,{\theta}}=\frac{\langle D_{{\theta^+}}\rangle-\langle D_{{\theta^-}}\rangle}{\Delta\theta}
\end{equation}
where $\langle D_{\theta^\pm}\rangle$ are the mean data vectors from the varying cosmologies for the parameter $\theta$, averaged over the $1500$ realizations. Eq. \eqref{eq:nuderi1} is used for all cosmological and primordial non-Gaussianity parameters but not $\sum m_\nu$, for which we use
\begin{equation}\label{eq:nuderi2}
    D_{,\sum m_\nu}=\frac{\langle D_{\sum m_\nu=0.4}\rangle-12\langle D_{\sum m_\nu=0.2}\rangle+32\langle D_{\sum m_\nu=0.1}\rangle-21\langle D_{\sum m_\nu=0}\rangle}{12(0.1)}
\end{equation}
instead, as advised in \cite{Villaescusa-Navarro:2019bje}. $\langle D_{\sum m_\nu=0}\rangle$ is measured from the fiducial simulations with Zeldovich initial conditions in consistency with the other quantities involving massive neutrinos. We plot the derivatives in Figure \ref{fig:deri}.

\section{On Choice of $N_{\rm bin}$ and Fisher Analysis Consistency Checks}\label{app:tests}
\begin{figure}[t]
    \centering
         \includegraphics[width=0.6\textwidth]{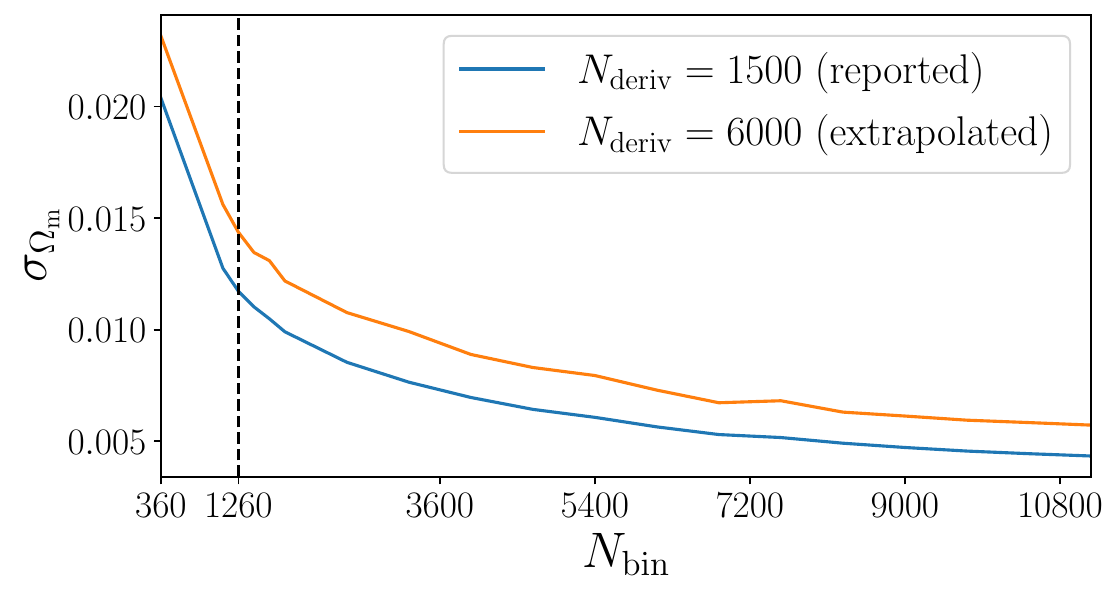}
        \caption{\textbf{Diminishing marginal return in information with $N_{\rm bin}$.} Here we show, taking $\Omega_{\rm m}$ as an example, that the $1$-$\sigma$ marginalized constraint tightens and eventually plateaus as we use more bins in our data vector. The blue line is (would be) the reported constraint from using all the existent $1500$ realizations in the estimation of the derivatives, while the orange marks the extrapolated constraint if $6000$ realizations were available, as explained in Appendix \ref{app:tests}. The dashed line marks our final choice of $N_{\rm bin}=1260$, which is conservatively decided to preserve the Gaussianity of the statistic and warrant that the constraints are reasonably converged.}
        \label{fig:convergencenbin}
\end{figure}

Unlike the power spectrum and the bispectrum, our statistic that supposedly probes a range of high-order correlators is not explicitly constrained by a low number of modes, and we may capture more information by having more bins in our data vector, until the point where the gain in signal is outweighed by the amount of noise introduced (Figure \ref{fig:convergencenbin}). In other words, increasing $N_{\rm bin}$ leads to tightened parameter constraints that are also generally less converged, given that the number of realizations available is limited. On the other hand, our way of estimating the Fisher matrix (Eq. \eqref{eq:Fisher}) requires that our statistic has a Gaussian likelihood. Since our data vector is built from feature counts, it will deviate from Gaussianity if the counts are low for a sufficiently large $N_{\rm bin}$.

\begin{figure}[t]
    \centering
         \includegraphics[width=1\textwidth]{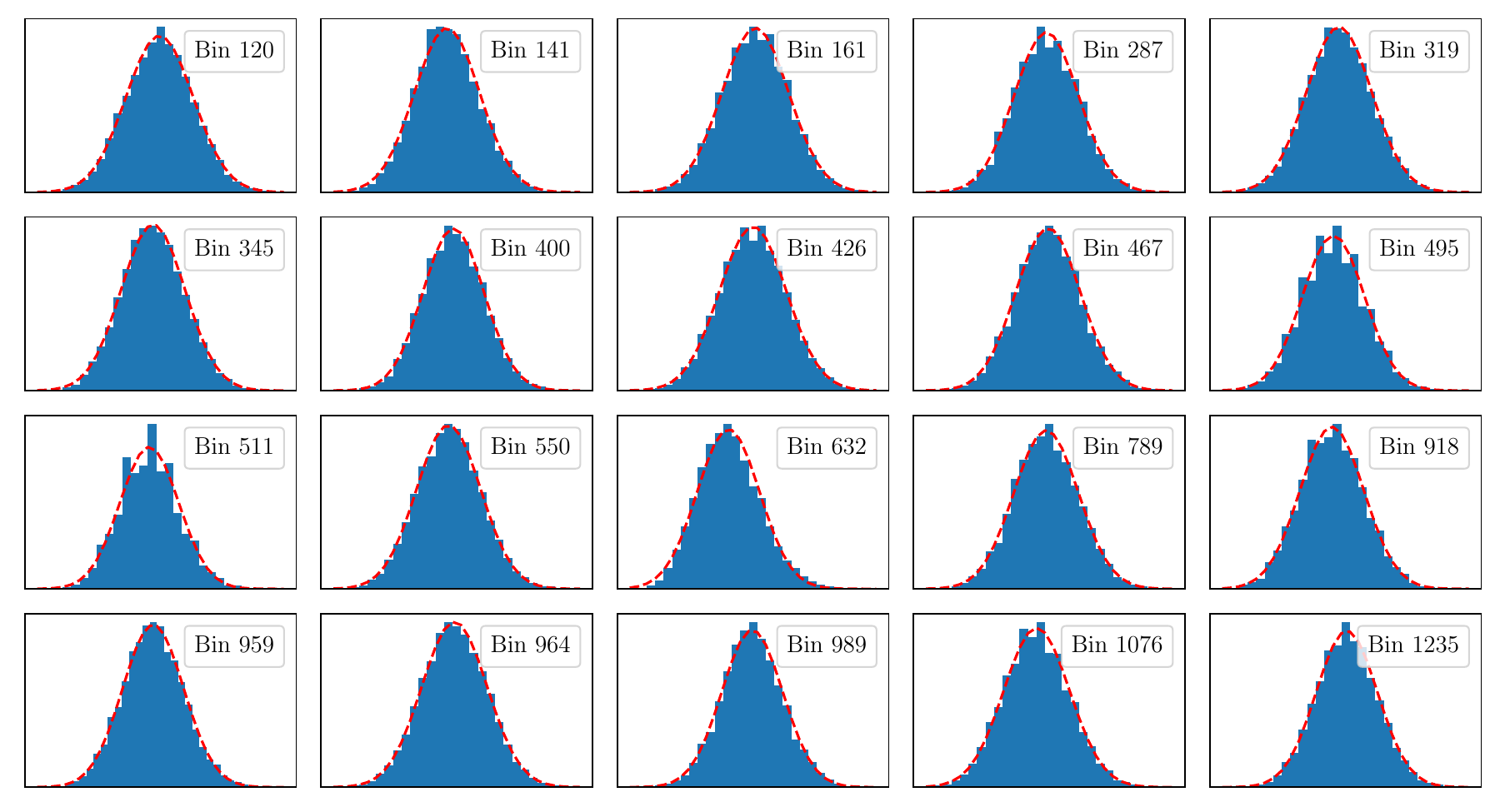}
        \caption{\textbf{Gaussianity of each bin.} Displayed here are histograms for $20$ bins, each randomly selected from the full fiducial data vector and containing $14500$ data points. Dashed lines in red are Gaussian fits. Visually there is no significant deviation from Gaussianity.}
        \label{fig:gauss}
\end{figure}

All things considered, we find $N_{\rm bin}=1260$ to be a suitable choice for making our point that persistent homology yields powerful statistics, as well as for trustworthy forecast estimates. We report on the following consistency checks:

\paragraph{Gaussianity.} We perform two tests:
\begin{enumerate}
    \item Kolmogorov-Smirnov test: This test checks the Gaussianity of each bin. We draw $50$ random realizations and average over $20$ draws. For all parameters, we find that $>96\%$ of all the $1260$ bins pass the test within the $95\%$ confidence interval. We also visually inspect the bins, and the histograms for $20$ randomly selected bins are presented in Figure \ref{fig:gauss}.
    \item Henze-Zirkler multivariate normality test \cite{henze:1990}: We implement this more general test that takes into account the Gaussianity across bins, using the \textsc{pingouin} library. Our data vector passes the test within the $95\%$ confidence interval for all parameters.
\end{enumerate} 

As previously mentioned, non-Gaussianity arises from low feature count, and in practice we may manually remove low-count bins to push for a larger $N_{\rm bin}$. However, we have chosen to maintain the simple, non-ad hoc construction of our data vector, as the aim of this work is not to fine-tune for the tightest possible Fisher constraints, but rather to provide a proof of concept for follow-up studies on using persistent homology for generic tasks.

\paragraph{Convergence.} We check the convergence of the numerical covariance and derivatives. Similar to \cite{Hahn:2019zob,Hou:2022rcd}, we calculate the ratio of $\sigma_\theta(N_{\rm sim})$, the $1$-$\sigma$ marginalized constraint on $\theta$ from $N_{\rm sim}$ realizations, to the constraint from all available realizations. Each convergence curve is the mean over $50$ random orderings of the realizations, and each error bar indicates one standard deviation above and below.

\begin{figure}[t]
    \centering
         \includegraphics[width=1\textwidth]{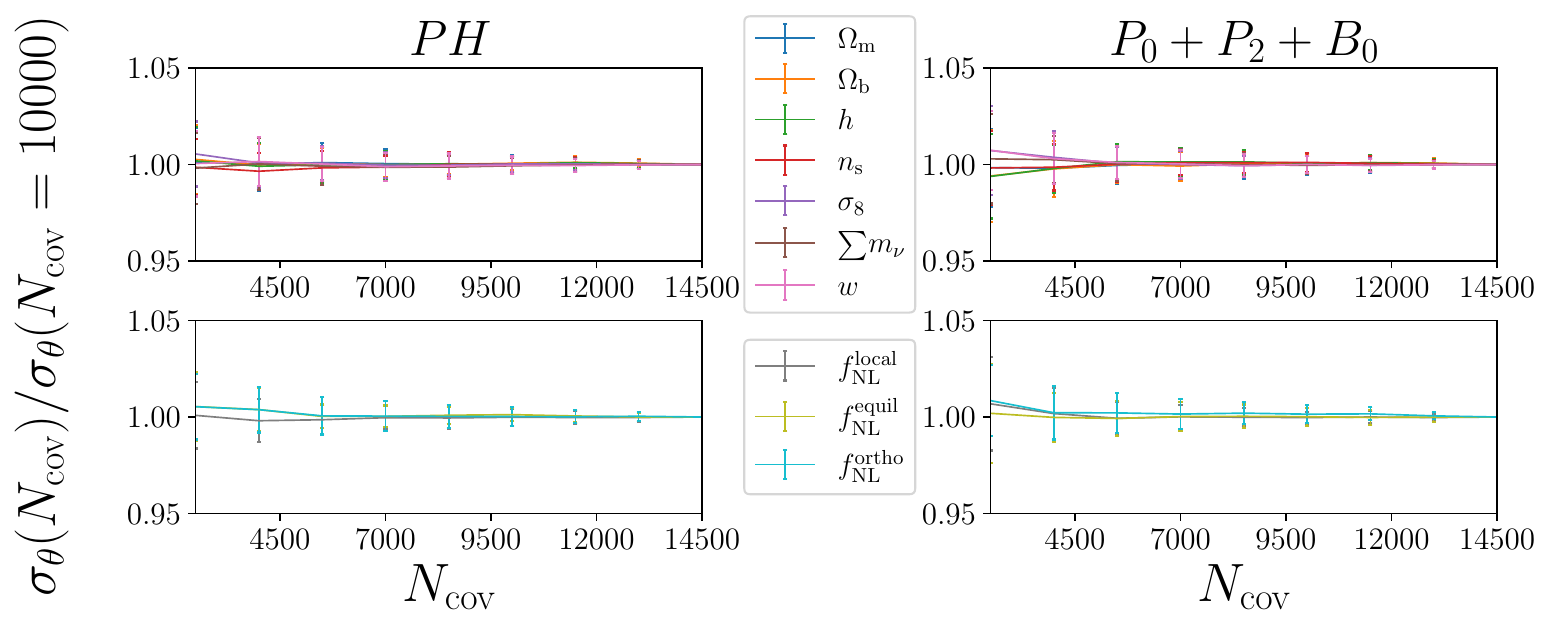}
         \caption{\textbf{Convergence of the numerical covariance.} \emph{Left Panel:} Our summary statistic from persistent homology. \emph{Right Panel:} Joint power spectrum and bispectrum statistic. The convergence curves are functions of the number of fiducial realizations $N_{\rm cov}$ used in estimating the covariance. Each error bar indicates one standard deviation above and below the mean from $50$ random orderings of the realizations. The fast convergence implies that the covariance is well estimated.}
         \label{fig:convergencecov}
\end{figure}

\begin{figure}[t]
    \centering
         \includegraphics[width=1\textwidth]{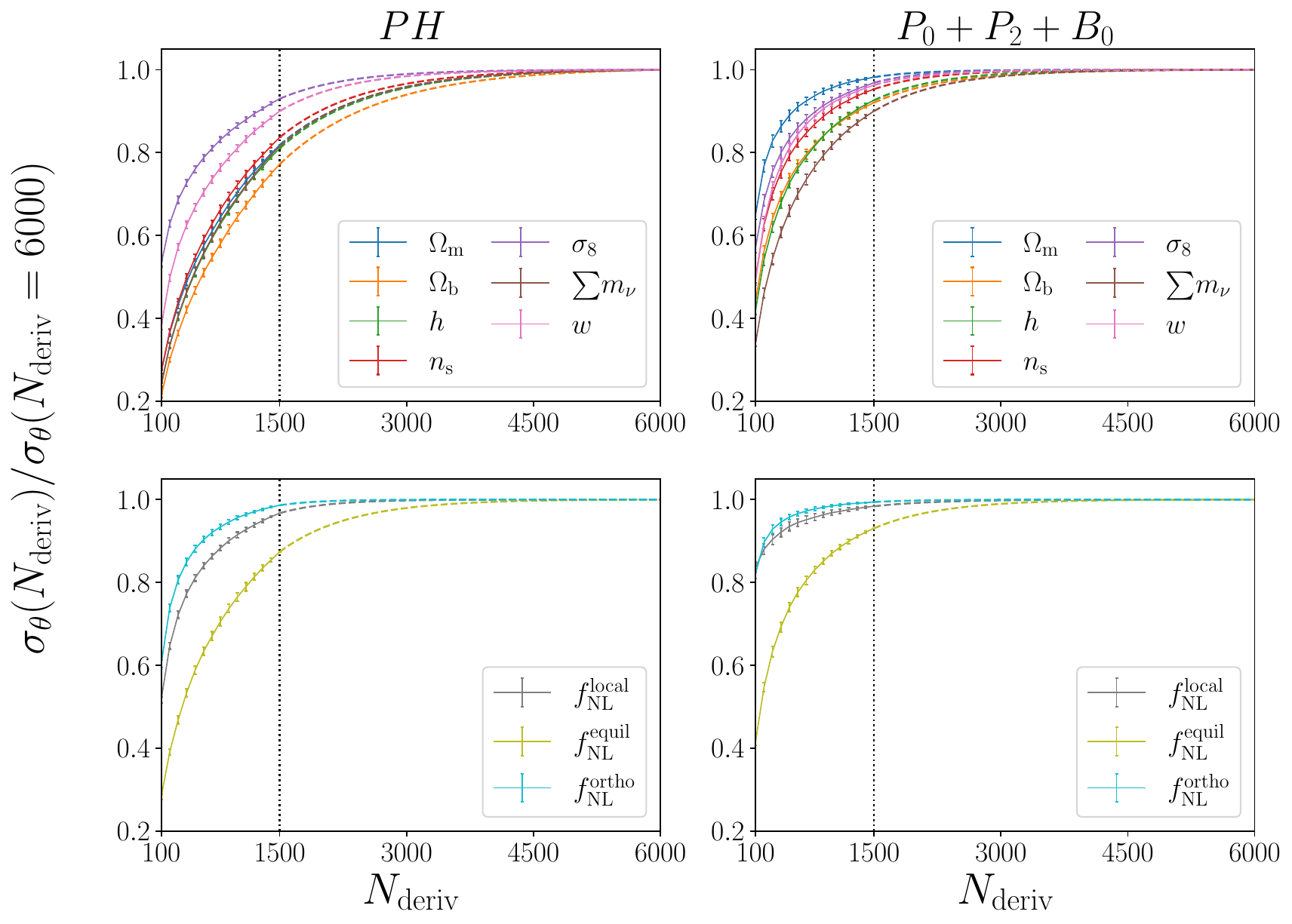}
        \caption{\textbf{Convergence of the numerical derivatives.} \emph{Left Panel:} Our summary statistic from persistent homology. \emph{Right Panel:} Joint power spectrum and bispectrum statistic. The convergence curves are functions of the number of realizations $N_{\rm deriv}$ used in estimating the derivatives. Each error bar indicates one standard deviation above and below the mean from $50$ random orderings of the realizations. Dashed colored lines are extrapolations using Eq. \eqref{eq:fit}. Vertical dotted lines in gray separate the existent realizations (to the left) from the hypothetical ones (to the right).}
        \label{fig:convergence}
\end{figure}

The estimation of the covariance converges very quickly with the number of fiducial realizations used, for both our statistic as well as the joint power spectrum and bispectrum statistic (Figure \ref{fig:convergencecov}).
On the other hand, as found in recent works on Fisher analysis testing new summary statistics \cite{Hahn:2019zob,Hahn:2020lou,Hou:2022rcd,Paillas:2022wob,Coulton:2022qbc,Coulton:2022rir,Jung:2023kjh}, the convergence of the derivatives is not ideal. In short, for half of the parameter set, it requires more simulations than we have available to moderately average out the numerical noise in the derivatives (Figure \ref{fig:convergence}). This issue is more thoroughly addressed in a recent paper \cite{Coulton:2023sfu}. In our case, we choose to estimate the uncertainty in our reported Fisher constraints by extrapolating the convergence curves to a large number of realizations, as done in \cite{Biagetti:2022qjl, Hou:2022rcd}. We employ the following parameterization
\begin{equation}\label{eq:fit}
    \sigma_\theta(N_{\rm sim}) = a_1 a_2^{-a_3 N_{\rm sim}}+a_4,
\end{equation}
where the constants $a_i$ are determined by fitting to each convergence curve using \texttt{optimize.curve\_fit} in the \textsc{scipy} library, taking into account the scatter around the mean from the $50$ orderings. The asymptotic value of a fit is then the predicted constraint given infinitely many realizations. We observe that the reported constraints from persistent homology are biased low and are expected to loosen by $\sim16\%$ on average. We apply the same treatment to the power spectrum and bispectrum and show the extrapolated constraints in Table \ref{tab:exresults}. The ``extrapolated result'' is identical to the reported result---constraints from persistent homology are tighter for all parameters except $\Omega_{\rm m}$ and $f_{\rm NL}^{\rm local}$. Hence, the conclusions of this work based on the reported constraints remain valid under the above consideration.

\begin{table}[t]
\begin{centering}
\begin{tabular}{ |P{2cm}|P{1cm}|P{2.7cm}|P{2.7cm}|P{2.7cm}|P{2.7cm}| } 
    \hline
    \multirow{2}{*}{$\theta$} & \multirow{2}{*}{\tiny{Fiducial}} & \multicolumn{2}{c|}{$N_{\rm deriv}=1500$ (reported)} & \multicolumn{2}{c|}{$N_{\rm deriv}=6000$ (extrapolated)} \\ \cline{3-6}
    & & \textcolor{ForestGreen}{$PH$} & \textcolor{RoyalBlue}{$P_{0}+P_{2}+B_{0}$} & \textcolor{ForestGreen}{$PH$} & \textcolor{RoyalBlue}{$P_{0}+P_{2}+B_{0}$} \\
    \hline
    \hline
        $\Omega_{\text{m}}$ & $0.3175$ & $\pm0.012$ & $\pm0.010$ & $\pm0.014$& $\pm0.010$\\
        $\Omega_{\text{b}}$ & $0.049$ & $\pm0.003$ & $\pm0.004$ & $\pm0.004$& $\pm0.004$\\
        $h$ & $0.6711$ & $\pm0.029$ & $\pm0.040$ & $\pm0.036$& $\pm0.043$\\
        $n_{\text{s}}$ & $0.9624$ & $\pm0.024$ & $\pm0.031$ & $\pm0.029$& $\pm0.033$\\
        $\sigma_8$ & $0.834$ & $\pm0.005$ & $\pm0.008$ & $\pm0.005$& $\pm0.009$\\
        $\sum m_\nu$ (eV) & $0$ & $\pm0.034$ & $\pm0.063$ & $\pm0.038$& $\pm0.065$\\
        $w$ & $-1$& $\pm0.051$ & $\pm0.101$ & $\pm0.062$& $\pm0.113$\\
    \hline
    \hline
        $f_{\text{NL}}^{\text{local}}$ & $0$& $\pm69$ & $\pm36$ & $\pm72$& $\pm37$\\
        $f_{\text{NL}}^{\text{equil}}$ & $0$& $\pm142$ & $\pm233$ & $\pm163$& $\pm250$\\
        $f_{\text{NL}}^{\text{ortho}}$ & $0$& $\pm61$ & $\pm70$ & $\pm62$& $\pm70$\\
    \hline
\end{tabular}
\par\end{centering}
\caption{\textbf{Extrapolated constraints.} The third and fourth columns are constraints from using all existent realizations, as reported in Table \ref{tab:results}. The rightmost two columns are the extrapolated constraints if $6000$ realizations were available. Both results are the same---constraints from persistent homology are tighter for all parameters except $\Omega_{\rm m}$ and $f_{\rm NL}^{\rm local}$. Hence, conclusions based on the reported constraints hold true under this consideration.}
\label{tab:exresults}
\end{table}

\section{Independent Information from Each Component of the Data Vector}\label{app:howinform}
As mentioned throughout the paper, in this work as a proof of concept, we do not fine-tune our setup for the tightest possible constraints, so the composition of our data vector should not be deemed optimal. Nevertheless, it is meaningful to gain an idea of how much non-overlapping information each part of the data vector contributes. We investigate this by computing the percentage increase in the $1$-$\sigma$ marginalized constraints from excluding various parts of the full data vector. A significant increase implies that the excluded part contains much independent information.

We consider three schemes: excluding a nearest-neighbor parameter $k$ value, excluding a homological dimension $p$, and excluding all birth $B_i$ or death $D_i$ histograms. The percentage increases are listed in Table \ref{tab:excludesigma}. We observe that all the $k$'s are similarly informative, with the exception of $k=1$ for $\sigma_8$ and $w$. On the other hand, the $0$-cycles are in general more informative than the $1$- and $2$-cycles, especially for $\sigma_8$ and $w$ (again). For birth and death histograms, they appear to be comparably important.

The percentages from excluding $k=1$ and $5$ back up our claim in the footnote\footref{fn:noalarm} under Section \ref{sec:results}, in the sense that, if we exclude $0$-cycles for those two $k$'s, constraints from persistent homology are still generally tighter, and our conclusions remain unchanged.

\begin{table}[t]
\begin{centering}
\begin{tabular}{|P{2cm}|P{0.8cm}|P{0.8cm}|P{0.8cm}|P{0.8cm}|P{0.8cm}|P{0.9cm}|P{0.8cm}|P{0.8cm}|P{0.8cm}|P{0.8cm}|P{0.8cm}|} 
    \hline
    \multirow{2}{*}{$\theta$} & \multicolumn{6}{c|}{\scriptsize{Exclude $k$}} & \multicolumn{3}{c|}{\scriptsize{Exclude $p$-cycles}} & \multicolumn{2}{c|}{\tiny{Exclude all $B_i$/$D_i$}} \\
    \cline{2-12} & \tiny{$k=1$} & \tiny{$k=5$} & \tiny{$k=15$} & \tiny{$k=30$} & \tiny{$k=60$} & \tiny{$k=100$} & \tiny{$p=0$} & \tiny{$p=1$} & \tiny{$p=2$} & \tiny{all $B_i$} & \tiny{all $D_i$} \\
    \hline
    \hline
        $\Omega_{\text{m}}$ & \scriptsize{$+13\%$} & \scriptsize{$+8\%$} & \scriptsize{$+9\%$} & \scriptsize{$+10\%$} & \scriptsize{$+8\%$} & \scriptsize{$+10\%$} & \scriptsize{$+33\%$} & \scriptsize{$+18\%$} & \scriptsize{$+22\%$} & \scriptsize{$+43\%$} & \scriptsize{$+34\%$}\\
        $\Omega_{\text{b}}$ & \scriptsize{$+8\%$} & \scriptsize{$+7\%$} & \scriptsize{$+11\%$} & \scriptsize{$+8\%$} & \scriptsize{$+10\%$} & \scriptsize{$+13\%$} & \scriptsize{$+23\%$} & \scriptsize{$+20\%$} & \scriptsize{$+23\%$} & \scriptsize{$+38\%$} & \scriptsize{$+40\%$}\\
        $h$ & \scriptsize{$+8\%$} & \scriptsize{$+11\%$} & \scriptsize{$+10\%$} & \scriptsize{$+14\%$} & \scriptsize{$+8\%$} & \scriptsize{$+11\%$} & \scriptsize{$+23\%$} & \scriptsize{$+24\%$} & \scriptsize{$+23\%$} & \scriptsize{$+49\%$} & \scriptsize{$+45\%$}\\
        $n_{\text{s}}$ & \scriptsize{$+11\%$} & \scriptsize{$+8\%$} & \scriptsize{$+8\%$} & \scriptsize{$+8\%$} & \scriptsize{$+7\%$} & \scriptsize{$+10\%$} & \scriptsize{$+34\%$} & \scriptsize{$+17\%$} & \scriptsize{$+17\%$} & \scriptsize{$+35\%$} & \scriptsize{$+38\%$}\\
        $\sigma_8$ & \scriptsize{$+35\%$} & \scriptsize{$+7\%$} & \scriptsize{$+4\%$} & \scriptsize{$+5\%$} & \scriptsize{$+4\%$} & \scriptsize{$+5\%$} & \scriptsize{$+81\%$} & \scriptsize{$+11\%$} & \scriptsize{$+10\%$} & \scriptsize{$+24\%$} & \scriptsize{$+22\%$}\\
        $\sum m_\nu$ (eV) & \scriptsize{$+8\%$} & \scriptsize{$+10\%$} & \scriptsize{$+10\%$} & \scriptsize{$+11\%$} & \scriptsize{$+12\%$} & \scriptsize{$+9\%$} & \scriptsize{$+28\%$} & \scriptsize{$+22\%$} & \scriptsize{$+19\%$} & \scriptsize{$+49\%$} & \scriptsize{$+51\%$}\\
        $w$ & \scriptsize{$+49\%$} & \scriptsize{$+7\%$} & \scriptsize{$+7\%$} & \scriptsize{$+7\%$} & \scriptsize{$+5\%$} & \scriptsize{$+6\%$} & \scriptsize{$+104\%$} & \scriptsize{$+13\%$} & \scriptsize{$+14\%$} & \scriptsize{$+29\%$} & \scriptsize{$+26\%$}\\
    \hline
    \hline
        $f_{\text{NL}}^{\text{local}}$ & \scriptsize{$+6\%$} & \scriptsize{$+4\%$} & \scriptsize{$+4\%$} & \scriptsize{$+4\%$} & \scriptsize{$+3\%$} & \scriptsize{$+5\%$} & \scriptsize{$+25\%$} & \scriptsize{$+9\%$} & \scriptsize{$+9\%$} & \scriptsize{$+16\%$} & \scriptsize{$+18\%$}\\
        $f_{\text{NL}}^{\text{equil}}$ & \scriptsize{$+8\%$} & \scriptsize{$+6\%$} & \scriptsize{$+7\%$} & \scriptsize{$+8\%$} & \scriptsize{$+9\%$} & \scriptsize{$+7\%$} & \scriptsize{$+22\%$} & \scriptsize{$+18\%$} & \scriptsize{$+17\%$} & \scriptsize{$+27\%$} & \scriptsize{$+26\%$}\\
        $f_{\text{NL}}^{\text{ortho}}$ & \scriptsize{$+11\%$} & \scriptsize{$+7\%$} & \scriptsize{$+5\%$} & \scriptsize{$+3\%$} & \scriptsize{$+2\%$} & \scriptsize{$+2\%$} & \scriptsize{$+45\%$} & \scriptsize{$+17\%$} & \scriptsize{$+8\%$} & \scriptsize{$+57\%$} & \scriptsize{$+60\%$}\\
    \hline
\end{tabular}
\par\end{centering}
\caption{\textbf{Excluding data vector parts.} Each percentage shown is the increase in the $1$-$\sigma$ marginalized constraint when a part of the data vector is excluded. A significant increase implies that the excluded part contains much independent information about that parameter. We consider three schemes: excluding a nearest-neighbor parameter $k$ value, excluding a homological dimension $p$, and excluding all birth $B_i$ or death $D_i$ histograms.}
\label{tab:excludesigma}
\end{table}

\section{On Applicability to the \sancho Suite of Galaxy Catalogs}\label{app:sancho}
The \sancho suite\footnote{\url{https://quijote-simulations.readthedocs.io/en/latest/sancho.html}, developed by Matteo Biagetti, Juan Calles, Jacky H. T. Yip, and Emilio Bellini.}, based on \quijote, contains a total of $240000$ galaxy catalogs at the fiducial and varying cosmologies primarily designed for Fisher analyses. \sancho allows for studying the Halo Occupation Distribution (HOD) model parameters \cite{Zheng:2007zg} in addition to the parameters surveyed in this work, and its integrity has been verified against a theoretical model based on the effective field theory of large-scale structure.

We have spent substantial time and efforts on applying our pipeline to \sancho. We could not trust the forecast, however, because our data vector from \sancho does not pass the Gaussianity tests: the distribution of data is bimodal for most bins. While the data vector can be made Gaussian by subsampling the galaxy catalogs, i.e., randomly removing galaxies such that the number density is uniform across catalogs, it happens that the data vector then responds to parameter change non-linearly (for the $\Delta\theta$ values inherited from \quijote), rendering the numerical derivatives unreliable.

Even though our statistic computed on \sancho cannot warrant a Fisher analysis because of the intractable likelihood that it seems to have, we emphasize that it can be used for inference tasks, e.g., training a neural network to map our statistic to cosmological parameters.

\clearpage
\section{Additional Plots}\label{app:addplot}
\begin{figure}[!htb]
    \centering
        \includegraphics[width=1\textwidth]{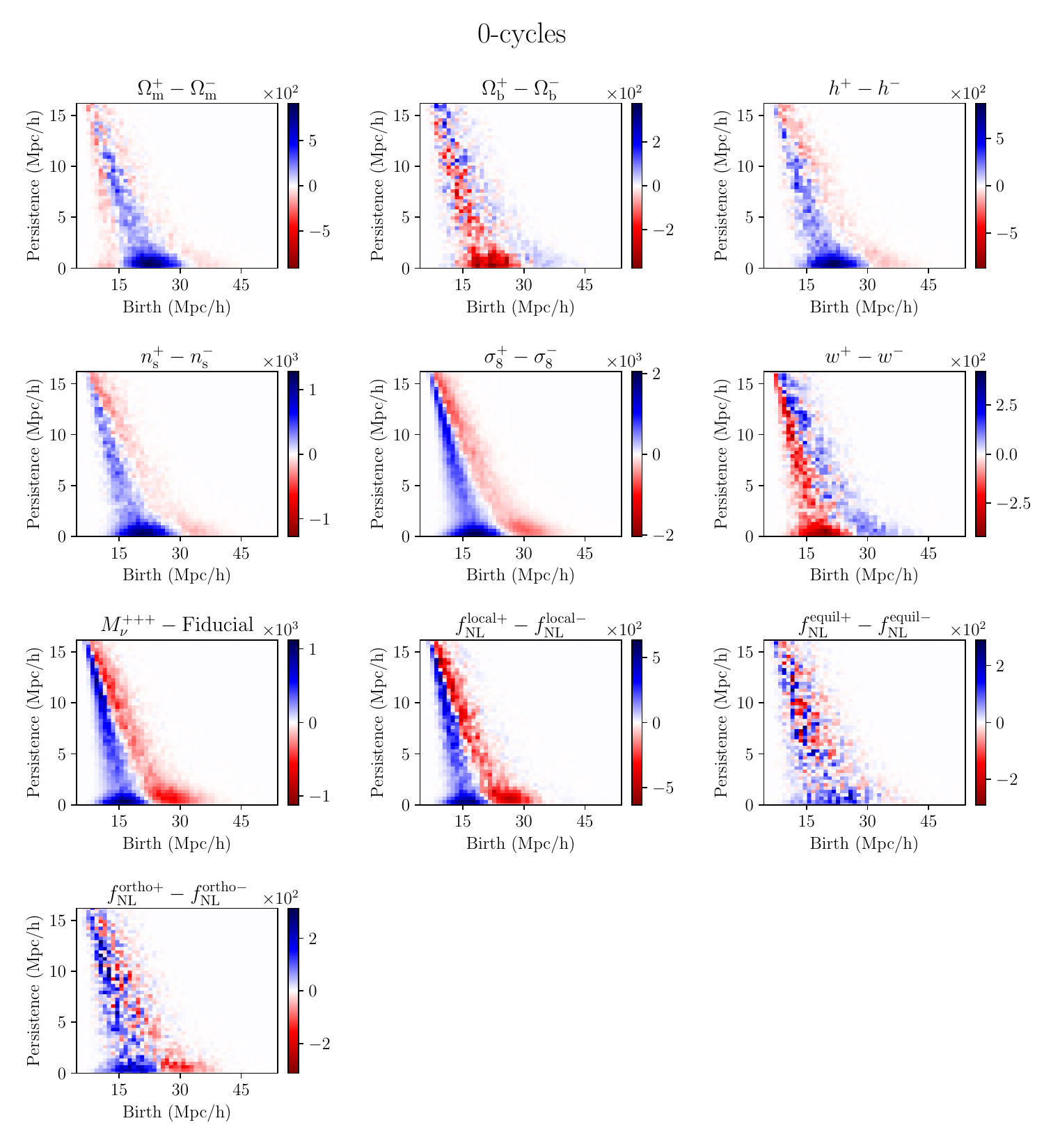}
        \caption{$0$-cycle mean persistence image differences for all parameters (averaged over $100$ realizations, $k=15$).} 
        \label{fig:pddiff0}
\end{figure}
\begin{figure}[!htb]
    \centering
        \includegraphics[width=1\textwidth]{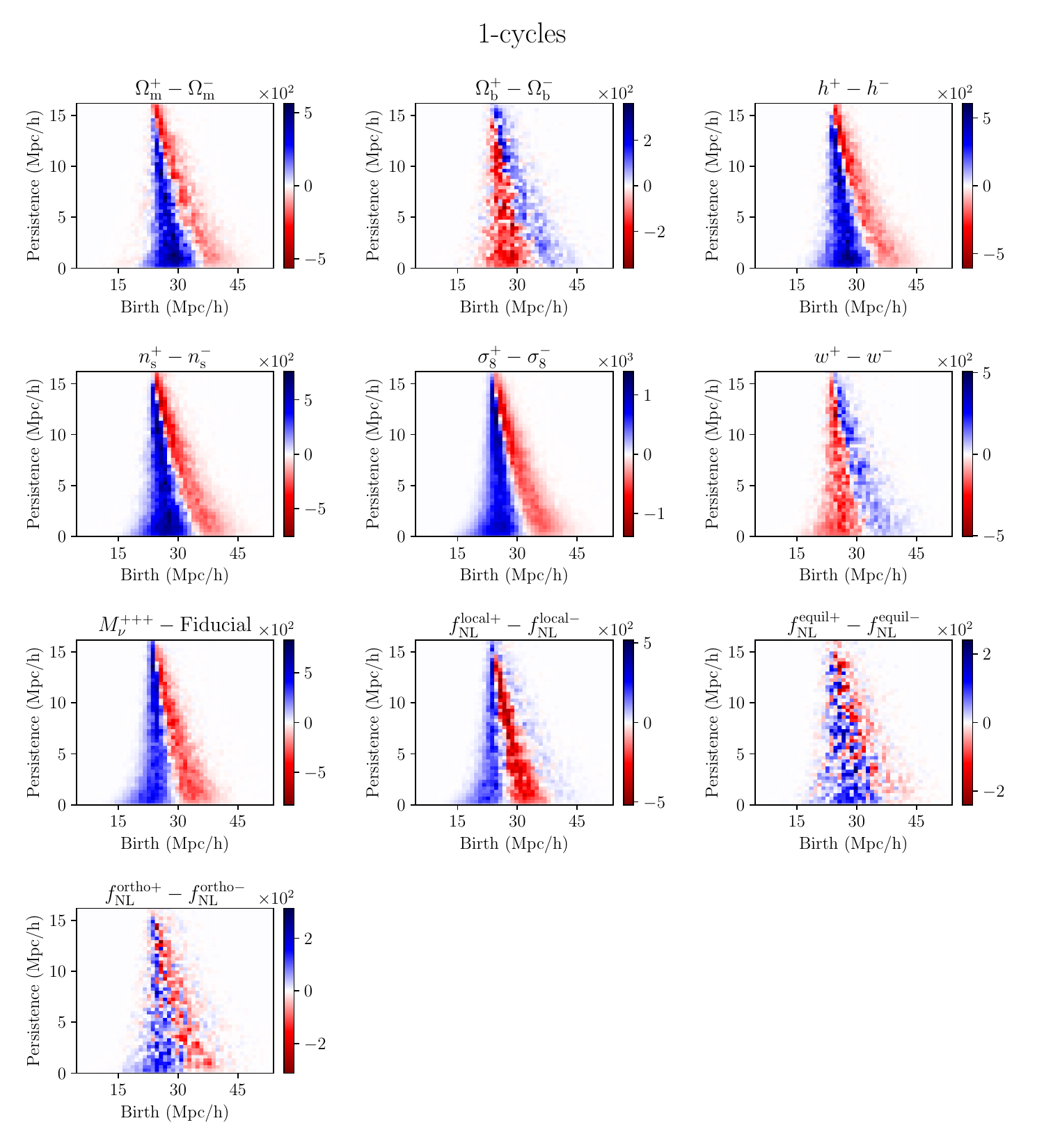}
        \caption{$1$-cycle mean persistence image differences for all parameters (averaged over $100$ realizations, $k=15$).} 
        \label{fig:pddiff1}
\end{figure}
\begin{figure}[!htb]
    \centering
        \includegraphics[width=1\textwidth]{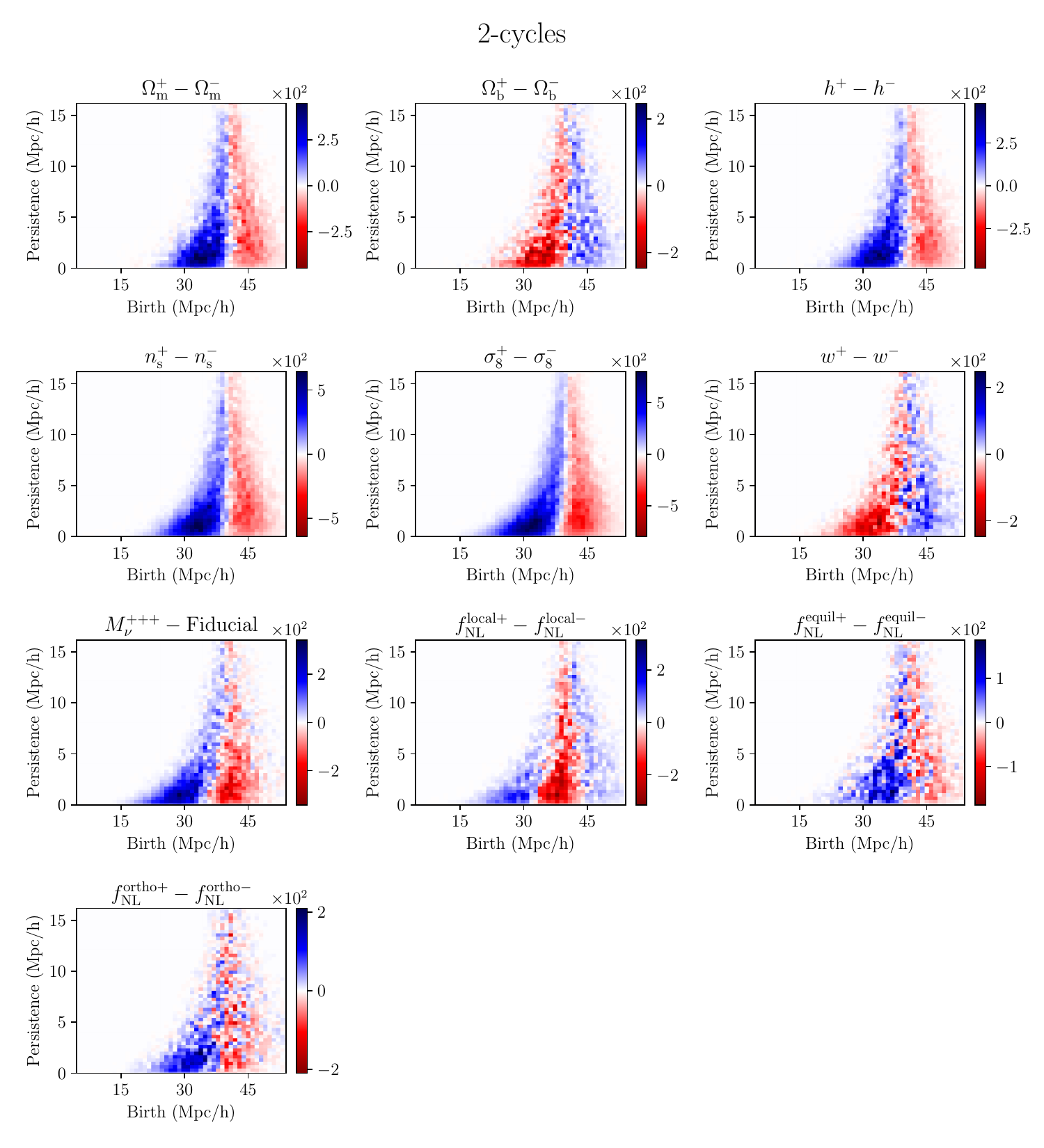}
        \caption{$2$-cycle mean persistence image differences for all parameters (averaged over $100$ realizations, $k=15$).} 
        \label{fig:pddiff2}
\end{figure}

\clearpage
\newpage
\bibliographystyle{utphys}
\bibliography{Halos-TDA}

\end{document}